\documentclass[aps,prb,twocolumn,showpacs,groupedaddress]{revtex4}
\usepackage[american]{babel}
\usepackage{color,graphicx,amsmath,amssymb,bm}

\newcommand\ket[1]{\left|#1\right\rangle}
\newcommand\bra[1]{\left\langle#1\right|}
\newcommand\avg[1]{\left\langle#1\right\rangle}
\newcommand\Avg[1]{\langle#1\rangle}
\newcommand\lavg{\left\langle}
\newcommand\ravg{\right\rangle}
\newcommand\be{\begin{equation}}
\newcommand\ee{\end{equation}}
\newcommand\la{\leftarrow}
\newcommand\ra{\rightarrow}
\newcommand\mbf{\mathbf}
\newcommand\mbb{\mathbb}
\newcommand\trm{\textrm}
\newcommand\id{{\rm 1} 
       \hspace{-1.1mm} {\rm I}
        \hspace{0.5mm}}
\newcommand\kbar{k\hspace{-2.2mm}\rule[1.75mm]{2mm}{0.5pt}}
\newcommand\bea{\begin{eqnarray}}
\newcommand\eea{\end{eqnarray}}
\newcommand\stocavg[1]{{\lavg #1 \ravg}_{\trm{stoc}}}
\newcommand\motavg[1]{{\lavg #1 \ravg}_{t}}
\newcommand\fk[1]{\mathfrak{#1}}
\newcommand\raw{\rightarrow}
\newcommand\law{\leftarrow}

\newcommand\eq[1]{Eq.~(\ref{#1})}
\newcommand\eqs[2]{Eq.~(\ref{#1}, \ref{#2})}
\newcommand\fig[1]{Fig.~\ref{#1}}

\begin{document}

\title{Transport properties of a periodically driven superconducting single electron transistor}
  
\author{Alessandro Romito}
\affiliation{
  Department of Condensed Matter Physics, 
  The Weizmann Institute of Science,     
  Rehovot 76100, Israel}
\affiliation{
  NEST-CNR-INFM \& Scuola Normale
  Superiore, Piazza dei Cavalieri 7, I-56126 Pisa, Italy}

\author{Simone Montangero}
\affiliation{NEST-CNR-INFM \& Scuola Normale
        Superiore, Piazza dei Cavalieri 7, I-56126 Pisa, Italy}

\author{Rosario Fazio}
\affiliation{International School for Advanced Studies (SISSA)
        via  Beirut 2-4,  I-34014, Trieste,  Italy}
\affiliation{NEST-CNR-INFM \& Scuola Normale
        Superiore, Piazza dei Cavalieri 7, I-56126 Pisa, Italy}

\date{\today}

\begin{abstract}

We discuss coherent transport  of Cooper pairs through a Cooper pair shuttle. We analyze both 
the DC and AC Josephson effect in the two limiting cases where the charging energy $E_C$ is either 
much larger or much smaller than the Josephson coupling $E_J$. In the limit $E_J \ll E_C$ we
present the detailed behavior of the critical current as a function of the damping rates and 
the dynamical phases. The AC effect in this regime is very sensitive to all dynamical scales present 
in the problem. The effect of fluctuations of the external periodic driving is discussed as well.
 In the opposite regime the system can be mapped onto the quantum kicked rotator, a classically 
chaotic system. We investigate the transport properties also in this regime showing that the 
underlying classical chaotic dynamics  emerges as an incoherent transfer of Cooper pairs through  
the shuttle. For an appropriate choice of the parameters the Cooper pair shuttle can exhibit 
the phenomenon of dynamical localization. We discuss in details the properties of the localized
regime as a function of the phase difference between the superconducting electrodes and the 
decoherence due to gate voltage fluctuations. Finally we point how dynamical localization is 
reflected in the noise properties of the shuttle.
\end{abstract}

\maketitle

\section{Introduction}
\label{intro}
Soon after the appearance of the microscopic theory of superconductivity~\cite{bardeen57}, 
Josephson predicted a remarkable manifestation of macroscopic quantum 
coherence~\cite{josephson62} by showing that two superconducting electrodes connected 
by an insulating barrier can sustain a dissipationless current.
Since its discovery, the Josephson effect has had a tremendous impact both in 
pure~\cite{tinkham96,barone82} and applied physics~\cite{barone82}. One of the most 
recent and exciting developments in the research based on the Josephson effect is 
probably in the implementation of superconducting nanocircuits for solid state quantum 
computation~\cite{makhlin01}. 

In nanodevices a new energy scale appears, the charging 
energy, and new interesting effects show up due to the interplay between Josephson 
coupling and the presence of charging. The Josephson coupling, leading to phase coherence 
between the two superconducting electrodes can be understood in terms of the coherent 
superposition of different charge states. Coulomb blockade~\cite{nato92} on the other side 
tends to localize the charge thus destroying phase coherence. The simplest example 
of this interplay is provided by behavior of the supercurrent through a Superconducting 
Single Electron Transistor (SSET)~\cite{tinkham96}. It consists in a small 
superconducting island connected, by tunnel junctions, to two superconducting electrodes.

Additional features emerge if the SSET is  coupled to mechanical degrees of freedom thus 
combining the field of single electron effects with the intensively studied area of  
Nano-ElectroMechanical systems~\cite{andreas96}. Among the most interesting devices in 
this area there is the electron shuttle (for a review see Ref.~\onlinecite{shekhter03}).  
In its essential realization, a shuttle system consists of a small conducting grain, in 
Coulomb blockade regime, oscillating periodically between two electrodes (source and drain).
The essential condition to characterize the shuttling mechanism is that the grain must 
be in contact with a single electrode at any time. Following the original proposal of 
a normal shuttle Gorelik {\it et al.}~\cite{gorelik01} introduce the {\em Cooper pair shuttle}
where all the device (electrodes and central island) is in the superconducting state.
Despite the fact that the central island is never connected to the two superconductors simultaneously, 
the Chalmers group has shown that the system is still capable 
to establish a global phase coherence and hence support a finite Josephson current~\cite{gorelik01,isacsson02}. 
The shuttle does not only carry charge, as in the normal metal case, but it also establishes
phase coherence between the superconductors. Differently from the normal metal case, 
the Cooper pair shuttle does not need a moving island, it  is just a SSET with time dependent 
Josephson couplings and therefore it can be realized in the standard SSET with time-dependent
fluxes~\cite{romito03}. The properties of the Cooper pair shuttle crucially depend on the 
decoherence mechanism  which is also responsible in driving the system toward a steady state.
The presence of dissipation modifies the current phase relation, but does not (in general) 
destroy the Josephson current~\cite{gorelik01,romito03}.
The effect of gate voltage fluctuations has been analyzed in Ref.~\onlinecite{romito03} where it has been 
shown that decoherence can even enhance the Josephson current. Additional work on the Cooper 
pair shuttle considered the full counting statistics of Cooper pair shuttling~\cite{romito04},  
and the possibility of observing quantum chaotic dynamics~\cite{romito05'}.

The present paper extends our previous works on the subject~\cite{romito03,romito05'}. 
In addition to our previous results we provide details of the derivation of 
DC Josephson current in the limit $E_J/E_C \ll 1$,
and consider several extensions which are important for a  connection with possible 
experiments. We also analyze the effect of fluctuations in the external driving and the effect of 
an external voltage. Moreover we discuss the AC effect and study the interplay of various 
times scales on the spectrum of the AC current. In the opposite limit  $E_J/E_C \gg 1$, 
which has not been discussed in the literature so far,
we present analytical and numerical results on the dynamical localization and discuss its 
signatures on the Josephson current fluctuations.

The paper is organized as follows. In Section~\ref{sistema} we present the model 
of the Cooper pair shuttle. In Section~\ref{calcoli} we analyze the transport properties 
of the shuttle in the Coulomb blockade regime. We present the details of the formalism to 
determine the steady state density matrix, and to derive the steady state Josephson current,
whose physical features are discussed in  Subsection~\ref{average_current}. 
Subsections~\ref{motion_fluctuation} and~\ref{ac}  are dedicated respectively to the effect 
of fluctuations of the external driving on the Josephson current and to the effect of an 
applied voltage bias, the AC Josephson effect. The chaotic regime of the Cooper pair shuttle is 
the subject of Section~\ref{chaotic}. We discuss the dynamics of the charge in the central island, 
in Subsection~\ref{JQKR_dynamics}. A new feature which appear as compared to the kicked rotator 
is an extra phase shift during the kicks which is due to the superconducting phase difference 
of the electrodes. This phase shift plays a key role since it is responsible for time reversal 
symmetry breaking whose consequences for the dynamics are investigated in Subsection~\ref{coetocue}.
In Sections~\ref{transport_chaotic} we develop the necessary formalism to calculate the 
full counting statistics in the chaotic regime. The concluding remarks are presented in 
Section~\ref{conclusions}. Several technical details are given in the 
Appendices~\ref{voltaggio_costante},~\ref{diffusion_corrections} and~\ref{levitov}. 
Throughout the paper  $k_B=1$.

\section{The model}
\label{sistema}

\begin{figure}[t!]
\begin{center}
\includegraphics[width=0.43\textwidth]{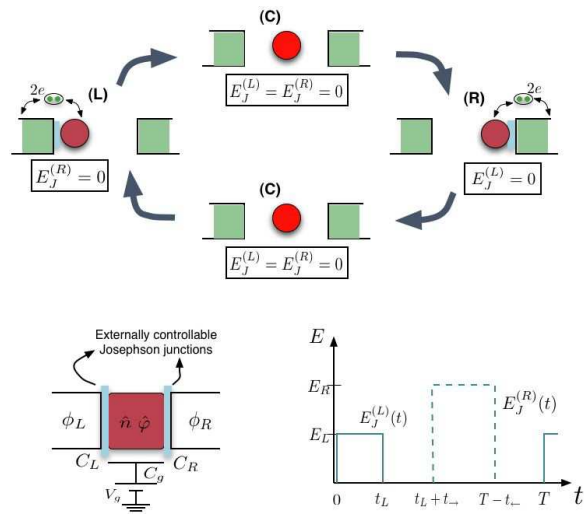} 
\end{center}
\caption{(Color online). Upper left panel. Schematic representation of the 
        system described 
        by the Hamiltonian \eq{h0}. It consists of a 
        Cooper pair box coupled through externally switched 
        Josephson junctions to phase biased 
        superconductors.
         Upper right panel. Time dependence of the left and 
        right Josephson energies within a single period.
        Lower panel. Sketch of the Cooper pair shuttle's cycle. 
        The three intervals L, C and R, within
        the period $T=t_L+t_{\rightarrow}+t_R+t_{\leftarrow}$, 
        correspond to the situations: (\textbf{L})
        ${E_J}^{(L)}(t) =E_L$, ${E_J}^{(R)}(t) =0$ 
        (interaction time at left lead); 
        (\textbf{C}) ${E_J}^{(L)}(t) =0$, ${E_J}^{(R)}(t) =0$ (free evolution
        time in forward and backward directions); 
        (\textbf{R}) ${E_J}^{(L)}(t) =0$, ${E_J}^{(R)}(t) =E_R$ 
        (interaction time at right lead). }
\label{evolution}
\end{figure}

The  Cooper pair shuttle is schematically shown in \fig{evolution}. It consists 
of a central island connected to two superconducting electrodes and capacitively
coupled to a gate voltage. The superconducting island is small enough such  that 
charging effects have to be included. The two leads are macroscopic and their
phases $\phi_{L,R}$ can be treated as classical variables. The couplings of the 
island to the leads are time-dependent. This time dependence is given by external 
means and can be achieved either by making the island to move or by tuning in 
time magnetic fluxes and gates. This has to be contrasted with the case of single
electron shuttle where for the implementation of the shuttle a mechanical moving 
island is necessary~\cite{erbe98,erbe01}.
 
The system is described by the following Hamiltonian
\begin{equation}
        H_0 = E_C(t) [\hat{n}- n_g(t)]^2 - \sum_{b =L,R} E_J^{(b)} (t)
        \cos(\hat{\varphi}-\phi_b) \, .
\label{h0}
\end{equation}
In Eq.(\ref{h0}) $\hat{n}$ is the number of excess Cooper pairs in the central 
island and $\hat{\varphi}$ is its conjugate phase, $[\hat{n}, \hat{\varphi} ]= -i$. 
The charging energy is given by $E_C(t)=(2e)^2/2C_{\Sigma}(t)$ with
$C_{\Sigma}(t)=C_g(t)+C_L(t)+C_R(t)$ the total capacitance of the SSET 
($C_{L/R/g}$ are the various capacitances indicated in Fig.~\ref{evolution}),
$E_J^{(L,R)}(t)$ are the Josephson couplings to the left or right lead respectively, 
and $n_g(t)=C_g(t) V_g(t)/2e$ is the gate charge which can be tuned via the  
gate voltage $V_g$.  

The Hamiltonian  in \eq{h0} is nothing else than a SSET with an external drive contained 
in the time dependence of the Josephson energies and of the gate voltage. 
If the time dependence of the coefficients is neglected the 
system is a SSET whose physics is known both in the case 
of macroscopic junctions ($E_J \gg E_C$) and in the presence of
charging effects ($E_J \ll E_C$)~\cite{tinkham96}.
By introducing a time dependence of the coefficients, it is possible to explore
different regimes. A case of adiabatic change of $E_J^{(b)}(t)$ is that 
of a Cooper pair sluice which has been experimentally and theoretically 
discussed in the literature~\cite{niskanen03,niskanen05}.
The shuttling mechanism we are interested in is essentially characterized by 
the sequence of time lapses during which the grain exchanges charges with the leads
and time intervals during which it is isolated from the leads.
The island is said to be in contact with one of the leads when  the corresponding 
Josephson coupling is non-zero (with value $E_L, \, E_R$) (configurations $L$ and 
$R$ in Fig.\ref{evolution}). In the intermediate region (configuration $C$), 
$E_J^{(L)}(t)=E_J^{(R)}(t)=0$. Note that both Josephson coupling are never on 
at the same time. As in Ref.~\onlinecite{isacsson02} we employ a sudden approximation
(which requires a  switching time $\Delta t \ll \hbar/E_{L(R)}$) and suppose $E_J^{(L,R)}(t)$ 
to be step functions in each region (see Fig.~\ref{evolution}). In case of 
a mechanical realization of the Cooper pair shuttle such an approximation 
is well justified due to the rapid decay of the Josephson coupling 
with the distance between the grain and the lead.
For later convenience we define the functions
\bea
        \Theta_{L}(t) &=& \theta(t)\theta(t_L-t) \, , 
\label{tL}\\ 
        \Theta_R(t) &=& \theta(t-(t_L+t_{\rightarrow}))\theta(t_L+t_{\rightarrow}+t_R-t) \, , 
\label{tR}
\eea 
in order to write the time dependent Josephson energies as 
\be
        E_J^{(b)}(t)=E_b \sum_{n \in \mathbb{N}}\Theta_{b}(t-nT) \,\,\, , 
        \,\,\,\,\, b \in \{ L,R \}
\label{te}
\ee
The total capacitance $C_{\Sigma}(t)$ is weakly dependent on time 
at the contact regions\footnote{If the Cooper pair shuttling is achieved
 by means of time-dependent fluxes, as described in 
 Subsection~\ref{fluxes}, $E_C(t)$ is a constant during the whole cycle.} and therefore we assume  
it to be constant during the intervals L and R (obviously the same hold for $E_C(t)=E_C$). 
In the intermediate region (C) it is not necessary to specify the exact time dependence of 
$E_C(t)$, as it will be clear in Section~\ref{calcoli}.

In the rest of the paper we study the transport properties of the Cooper pair shuttle.  
The transfer of charge is expressed by the presence of a current at left and right 
contacts. The corresponding current operators are, in the Schr\"odinger picture,
\bea
        \hat{I}_L(t) & = &
        2e  \frac{E_L}{\hbar}  \sin \left( \hat{\varphi} -\varphi_L \right) \, \Theta_L (t) \, ,
\label{current_L}\\
        \hat{I}_R(t) & = &
        2e  \frac{E_R}{\hbar}  \sin \left( \hat{\varphi} -\varphi_R \right) \, \Theta_R (t) \, ,
\label{current_R}
\eea
corresponding to the coherent exchange of  Cooper pairs between the grain and the 
left or right lead respectively. Due to the periodical external driving, any interaction 
with the external environment leads to a steady state, where every observable is periodic. 
We will essentially ignore transient effects and concentrate on the stationary values of
physical observables.

\subsection{Cooper pair shuttle with time-dependent fluxes}
\label{fluxes}

\begin{figure}[t]
\begin{center}
\includegraphics[width=0.45\textwidth]{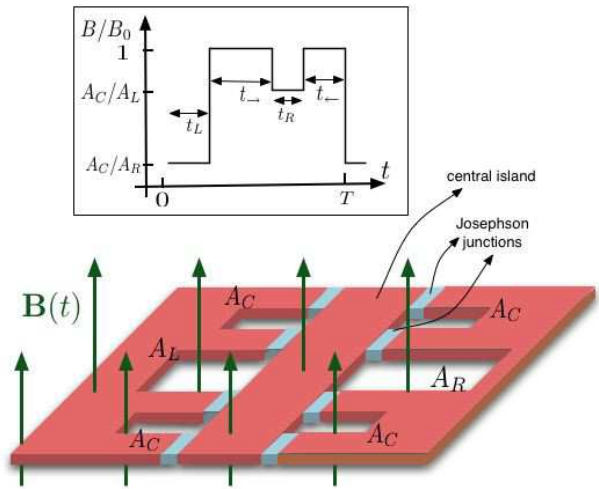}
\end{center}
\caption{(Color online). The setup for the 
implementation of the shuttle process
by means of a time-dependent magnetic field.
The inset shows the time dependence of the applied field
(in unity of $B_0=\Phi_0/(2A_C)$,$\Phi_0$ is the flux quantum) in order to realize Cooper 
pair shuttling. The different loop areas can be chosen in order to
obtain $E_L=E_R$.} 
\label{squid}
\end{figure}

Before analyzing in detail the transport properties, we  discuss a way 
to realize a Cooper pair shuttle which does not require any mechanically moving part. 
Here the time dependence of the Josephson couplings and $n_g$ is obtained by a time 
dependent magnetic field and gate voltage, respectively. The setup consists of a 
more complicated superconducting nanocircuit in a  uniform magnetic field as 
sketched in Fig.\ref{squid}. By substituting each Josephson junction by SQUIDs, 
it is possible to control the $E_J^{(b)}(t)$ by tuning the applied magnetic field piercing
the loop. The presence of three type of loops with different area, $A_L,A_R,A_C$ 
allows to achieve independently the three cases, where one of the two $E_J$'s is 
zero (regions L,R) or both of them are zero (region C), by means of a \emph{uniform} 
magnetic field. If the applied field is such that a half-flux quantum pierces the
areas $A_L$,$A_R$ or $A_C$, the Josephson couplings will be those of regions R,L and C, 
respectively and the Hamiltonian of the system can be exactly mapped onto that of Eq.(\ref{h0}).
Moreover, by choosing the ratios $A_C/A_{R}=0.146$, and $A_C/A_L=0.292$  the two Josephson 
coupling are equal, $E_L=E_R=E_J$. This implementation has several advantages. 
It allows to control the coupling with the environment by simply varying the time 
dependence of the applied magnetic field. The time scale for the variation of the magnetic 
field should be controlled at the same level as it is done in the implementation of 
Josephson nanocircuits for quantum computation (see Ref.~\onlinecite{makhlin01} for an 
extensive discussion). 

For a quantitative comparison with the results described here, the magnetic field should 
vary on a time scale shorter than $\hbar / E_J$, typically a few nanoseconds with the parameters of
Ref.~\onlinecite{nakamura99}. This is possible with present day technology~\cite{buisson03}.
At a qualitative level the features of the Josephson current presented in this paper do not 
rely on the step-change approximation of the Josephson couplings. 
These effects are observable even if the magnetic field changes on time-scales comparable 
or slower than $E_J$. The only strict requirement is that only one Josephson coupling at the 
time is switched on.
 
\section{Coulomb blockade regime}
\label{calcoli}

We first consider the system when $E_L, E_R \ll E_C$, {\it i.e.} 
in the Coulomb blockade regime. In addition, the gate voltage is chosen so that 
$0 < n_g(t) \ge 1/2$.  Namely  $n_g(t)=1/2$ when the system is in contact with one of the lead
and $n_g(t)=\textrm{const.} \in (0,1/2)$ in the remaining time of the cycle.
Our choice (the same of Ref.~\onlinecite{gorelik01}) results in having exact charge 
degeneracy during the Josephson contacts, then enhancing charge transfer. A 
different condition, of easier experimental realization, in which
 $n_g(t)=\textrm{const.}$  is discussed in Appendix \ref{voltaggio_costante}. 
In this limit one can restrict the Hilbert space of the system 
to the one spanned by the two charge states 
$\{\ket{n=0}, \ket{n=1}\}$. The Hamiltonian of the system restricted to the 
two dimensional vector space, reads
\be
        \hat{H}_0= E_C(t)[\frac{1}{2}-n_g(t)]\sigma_z - \! \sum_{b=L,R} 
        \! \frac{E_J^{(b)}(t)}{2} 
        \left( e^{-i \phi_b} \sigma_+ + \mbox{h.c.}  \right)
        \label{restricted_h0}
\ee 
where we used the $2 \times 2$ Pauli matrices $\sigma_i$ ($i=x,y,z$) with the standard 
notation $\sigma_{\pm}=(\sigma_x \pm \sigma_y)/2$.

In order to evaluate  the current, Eqs.(\ref{current_L},\ref{current_R}), 
or the average value of any observable, we need to compute the reduced 
density matrix of the central island $\rho (t)$. 
The steady state density matrix  depends on the specific decoherence mechanism.
The main source of decoherence in the Cooper pair shuttle is due to gate voltage fluctuations,
either induced by the electromagnetic environment or by background charges.

\subsection{Classical noise}

At a classical level voltage fluctuations can be included by adding a classical 
stochastic term to $n_g(t)$. The Hamiltonian in Eq.~(\ref{restricted_h0}) is modified
by the presence of the extra term,
\be
        \hat{H}=\hat{H}_0 +\xi(t) \sigma_z \, , 
\label{h_classical_noise}
\ee
where $\xi(t)$ has a white noise spectrum
\begin{eqnarray}
\stocavg{\xi(t)}& = &0 \nonumber \\
\stocavg{\xi(t)\xi(t')} & = & \hbar^2 \gamma \delta(t-t') \nonumber
\end{eqnarray}
where $\gamma$ is  the inverse decoherence time.
If we neglected the fluctuations, the time evolution of the system
would be fully coherent. By including fluctuations, the shuttle will be described
by $2 \times 2$ density matrix that obeys the following Bloch equation:
\begin{equation}
        \frac{\partial{\hat\rho}}{\partial t}=
        -\frac{\imath}{\hbar}
        \left( \hat H_0(t) \hat\rho - \hat\rho\hat H_0(t)\right) 
        -2 \gamma \left( \hat\rho - \sigma_z \hat\rho \sigma _z \right) \, .
\label{Bloch}
\end{equation}
The only stationary solution of this equation is trivial: 
$\hat\rho = \hat 1 /2$, this corresponds to the absence of any average 
superconducting current. This is a combined effect of the decoherence term and 
Josephson coupling. In the absence of Josephson coupling, voltage 
fluctuations can not cause transitions between the charge states 
so no relaxation takes place. With Josephson coupling switched on, 
the voltage fluctuations cause transitions between  the stationary states separated
by energy $E_J^{(L,R)}$. Classical voltage fluctuations result in equal transition 
rates with increasing and decreasing energy. The vanishing of the critical 
current has a simple explanation, the classical noise mimics a bath at high
temperature~\footnote{The high temperature limit is intended $T_b \gg E_J$, but, 
in any case, $T_b \ll E_C$, such that we can limit our discussion to two charge states.}. 
No coherence can be established at temperatures
much higher than the Josephson coupling energy. Nevertheless this model is 
worth to be considered because, as shown in Ref.~\onlinecite{romito04} there is an high 
temperature regime where the average current is zero but still coherence 
manifests in the higher moments of current fluctuations.
At low temperatures, $T_b \lesssim E_J^{(L,R)}$ the interactions with the 
bath can lead to a density matrix $\hat\rho \ne \hat 1$ and then to a non-vanishing 
supercurrent.

\subsection{Quantum noise}
 
In order to analyze the low temperature regime, we need to take into account the quantum 
features of the bath. As the most important source of fluctuations in the charge regime 
are gate voltage fluctuation we couple the shuttle via the charge operator $\hat{n}$ to 
an environment described by the Caldeira--Leggett model~\cite{weiss99},
\begin{equation}
        H_{int} = \hat{n}\hat{\mathcal{O}}+H_{bath}=
        \hat{n} \sum_i \lambda_i (a_i+ a_i^{\dag})+H_{bath} \; .
\label{hi}
\end{equation}
In Eq.(\ref{hi}), $H_{bath}$ is the bath Hamiltonian, with boson
annihilation/creation operators of the $i-$th mode $a_i$, $a_i^{\dag}$,
$H_{bath}=\sum_i \omega_i (a_i^{\dag}a_i + 1/2)$.
Due to the periodicity of the external driving the time evolution of the 
system at long time, $t \gg T$, can be determined by iterating the evolution of 
the density matrix $\rho(t)$ over one single period. This evolution  can be 
computed through a linear map $\mathcal{M}_{t\raw t+T}$ defined by 
\be
        \rho (t+T) = \mathcal{M}_{t\ra t+T} \left[\rho (t) \right].
\label{map}
\ee
With the following choice of parametrizing $\rho(t) = 1/2 \left[ {\rm 1} 
\hspace{-1.1mm} {\rm I}\hspace{0.5mm} +\mathbf{\sigma} \cdot \mathbf{r}(t) \right]$,
where $i=x,y,z$ and $r_i(t)=\lavg \sigma_i \ravg$, the map in Eq.~(\ref{map}) 
assumes the form of of a general affine map for the vector $\mathbf{r} (t)$:
\be
        \mathbf{r}(t+T)=M_t \mbf{r}(t)+\mbf{v}_t \, , \,\, 
        \mathbf{r} \in \mathcal{B}_1 (\mathbf{0}) \subset \mathbb{R}^3
\label{last_name}
\ee
where $\mathcal{B}_1$ is the Ball of unitary radius in $\mathbb{R}^3$.
The matrix $M_t$ fulfills the property 
\be
        \left| M_t \mbf{v} \right| \leq \left|\mbf{v} \right| \,
        \forall \mbf{v} \in \mathcal{B}_1(\mathbf{0})\, ,
\label{property}
\ee
as we will see from its explicit form determined below.

In the long time limit, the system reaches a periodic steady state,
\be
        \mbf{r}_{\infty}(t)=\left( \id-M_t \right) \mbf{v}_t \, ,
\label{steady}
\ee
if, and only if, $\det (\id-M_t) \neq 0$. When this condition is not satisfied 
the external bath  introduced in \eq{hi} is not effective and the system never 
loses memory of the initial conditions.

The stationary limit is the fixed point of $\mathcal{M}_{t\raw t+T}$\cite{shytov01}.
The expression of $\mbf{r}_{\infty}(t)$, and therefore of $\rho_{\infty}(t) = 1/2 \left[ {\rm 1} 
\hspace{-1.1mm} {\rm I} \hspace{0.5mm} +\mathbf{\sigma} \cdot \mathbf{r}_{\infty}(t) \right] $,
uniquely determines the steady state of the system. 

The periodic time dependence of any  physical observable $A$ is given by
\be
        \Avg{\hat{A}(t)} = \mathop{Tr}\{ \rho_{\infty}(t) \hat{A} \} \, ,
\ee
where the operator $\hat{A}$ is in the Schr\"odinger representation.

The assumption of a stepwise varying Hamiltonian considerably simplifies the form of 
the map $\mathcal{M}_{t \raw t+T}$, which now can be expressed as a composition of the 
time evolutions of $\rho$ in the intervals L, C, R (see Fig.\ref{evolution}). 
In each time interval it is straightforward to solve the corresponding master equation 
for the reduced density matrix~\cite{cohen}. In the portion of the cycle corresponding to 
the island being in contact with the left electrode the  master equation reads
\begin{equation}
        \dot{\mathbf{r}}(t) = G_{L} (t) \mathbf{r}(t) + 
        2 \gamma_{L} \mathbf{w}_{L} 
\label{master}
\end{equation}
with $\mathbf{w}^{\dag}_{L} = \tanh (E_{L} /T_b) \begin{pmatrix} \cos \phi_{L},&
\!\sin\phi_{L}, & \! 0 \end{pmatrix}$
and
\begin{equation}
        G_{L} =
        \begin{pmatrix}
        -2 \gamma_{L} & 0 & -\frac{E_{L}}{\hbar} \sin\phi_{L} \cr 0 & -2
        \gamma_{L} & -\frac{E_{L}}{\hbar} \cos\phi_{L} 
        \cr \frac{E_{L}}{\hbar} \sin\phi_{L} & \frac{E_{L}}{\hbar} \cos\phi_{L} & 0
\end{pmatrix} \; .
\end{equation}
Here, $\gamma_{L}$ is the dephasing rate (for this portion of the cycle), depending 
on the temperature of the bath, which is taken in thermal equilibrium at temperature $T_b$. 
The master equation when the island is in contact with the right superconducting lead 
goes along the same lines with the substitution $L \rightarrow R$ (thus introducing  
a dephasing rate $\gamma_{R}$). Both dephasing rate can be obtained 
in the Born-Markov approximation~\cite{cohen}, which requires that the bath 
autocorrelation time is the smallest  time scale in the problem. This treatment is valid
provided that $\gamma_{L(R)} \ll T_b/\hbar, E_{L(R)}/\hbar$, and that the time
interval $t_{L(R)}$ is much longer than both $\hbar T_b^{-1}$ and $\hbar E_{L(R)}^{-1}$. 
As an example, for an Ohmic bath with coupling to the environment  $\alpha \ll 1$, one has 
$\gamma_{L(R)}= (\pi/2) \alpha E_{L(R)} \coth (E_{L(R)}/2T_b)/\hbar$~\cite{weiss99}.
$G_L$ is time independent as a consequence of the Born-Markov approximation.
The solution of the  master equation in the contact region can be obtained in the form
\be
\mathbf{r}(t_L)=\exp \left( G_L t_L \right) \mathbf{r}(0) - 2 \gamma_L G_L^{-1} \cdot
\left[ \id - \exp \left( G_L t_L \right) \right]\mathbf{w}_L \, .
\label{solution} 
\ee
The parameters of the Hamiltonian enter the final results only through the combinations 
$\theta_{L(R)}=E_{L(R)} t_{L(R)}/\hbar$ and $\gamma_{L(R)} t_{L(R)}$.  Due to the condition 
$\gamma_{L(R)} \ll E_{L(R)}/\hbar$ the parameter $\hbar \gamma_{L(R)}/E_{L(R)}$ does not enter the 
results at lowest order.

During that part of the cycle when the island is disconnected from both electrodes, 
the situation is simpler. Since $\hat{n}$ is conserved, the evolution can be determined 
exactly
\begin{equation}
        \mbf{r}(t_{\raw}+t_L) = \exp(G_{\raw}t_{\raw}) 
        \cdot R(\chi_{\raw}) \mbf{r}(t_L) \,\, .
\label{central}
\end{equation}
In the previous equation we defined
\begin{equation}
        G_{\raw} =\begin{pmatrix}
        -\gamma_{\ra} & 0 & 0 \cr
        0 &  - \gamma_{\ra}  & 0 \cr
        0 & 0 & 1
        \end{pmatrix} 
\end{equation}
and
\begin{equation}
        R(\chi_{\ra}) = \begin{pmatrix}
         \cos(\chi_{\ra}) & -\sin(\chi_{\ra}) & 0 \cr
         \sin(\chi_{\ra}) &  \cos(\chi_{\ra}) & 0 \cr
        0 & 0 & 1
\end{pmatrix} . 
\end{equation}
where $\chi_{\ra} = \int_{t_L}^{t_L+t_{\ra}} E_C(t) (1-2n_g)/\hbar$. 
The rate $\gamma_{\ra}$ depends only on the properties of the bath. Its explicit 
time-dependence varies when the time scale is compared with the inverse ultraviolet bath mode 
cut-off, $1/\omega_c$, and the inverse bath temperature, $\hbar/T_b$~\cite{palma97,reina02}.
An expression of $\gamma_{\ra}$ in terms of bath parameters can be obtained within 
the same Born-Markov approximation discussed above in the case of a weakly coupling between 
the bath and the system. It gives $\gamma_{\ra}= 2 \pi \alpha T_b/\hbar$ in which case 
$\gamma_{\ra}$ is independent on time and the decay is purely exponential. 
The same equation holds in the backward free evolution time 
\be
         \mbf{r}(T) = \exp(G_{\law}t_{\law}) 
         \cdot R(\chi_{\law}) \mbf{r}(T-t_{\law}) \, ,
\ee
where $G_{\law}$ is defined as $G_{\raw}$ in \eq{central} with the replacement 
$\gamma_{\raw} \raw \gamma_{\law}$. In addition to the dynamical phases $\chi_{\ra(\la)}$ and 
$\theta_{L(R)}$, also the phase difference $\phi = \phi_L -\phi_R$  enters in  determining the  
physical observables. The effect of damping is characterized by the dimensionless quantities 
$\gamma_{L(R)} t_{L(R)}$, and $\gamma_{\ra(\la)} t_{\ra(\la)}$. 

From Eqs.~(\ref{master}-\ref{central}) it is easy to check that  $M_t$ fulfills the property 
in Eq.~\ref{property} except for the following values   
$(\gamma_L,\gamma_R,\gamma_{\ra},\gamma_{\la})=(0,0,0,0)$ or
$(\gamma_L,\gamma_R,\theta_{L},\theta_{R})=(0,0,k \pi/2,h \pi/2)$), 
$k, \, h$ integers, when  $\det(\id-M_t)=0$. In these cases, the system keeps memory of
its initial conditions and it never approaches the steady state.
This is however an artificial situation, because other sources of dissipation 
are present and will drive the system  to a steady state.

Let us comment on the assumption of Heaviside functions for $E_{L(R)}(t)$ we used to determine the steady state density matrix. 
It defines the simplest model to catch the features of the shuttling mechanism, i.e. 
the existence of different regimes during a single period time evolution.
In fact the precise shape of the Josephson energy pulses is not relevant, changing 
it will change the definition of the dynamics phases $\chi$, $\theta$, -eqs. (18), (21)- 
which enter as parameters in the results for the density matrix.
What is indeed neglected in our model is the effect of exciting higher energy modes and the 
effects of gate voltage fluctuations during the switching time. These are good approximations for switching times $\Delta t \lesssim \hbar/E_J$  and $\Delta t \gamma_i \ll 1$ for any dephasing rate $i \in \{ \rightarrow, \leftarrow, L, R\}$.

\subsection{DC Josephson current}
\label{average_current}

The asymmetry between emission and absorption of quanta from the bath leads to a nontrivial 
fixed point (\eq{steady}) for the map $\mathcal{M}$, thus leading to a non vanishing 
Josephson current through the Cooper pair shuttle. The current depends on the  quantum dynamical 
evolution of the charge on the island and on the interplay between the decoherence and 
the periodic driving.  If, for example, the period T is much larger than the inverse dephasing 
rates, the shuttle mechanism is expected to be inefficient and the critical
current is strongly suppressed. In the following we will describe a quite rich scenario, 
depending on the relative value of the various time scales and phase shifts.

As charge is conserved by the coupling to the environment, $[\hat{n},\hat{H}_{int}]=0$, 
current can flow only trough the electrodes. Therefore, in the Heisenberg picture,
$ \hat{I}_L(t) +  \hat{I}_R(t) + \dot{\hat{n}} =0$. By integrating over a period the 
average current reads ($I_R = - I_L \equiv I$)
$$
         I =\mbox{Tr} 
        \{ \hat{n} (\rho(t_L)- \rho(0)) \} = 
        \mbox{Tr} \{ \hat{n} (\rho(T/2)- \rho(0)) \} \,\, .
$$
We set the initial time within a period at the beginning of contact 
with the left lead.

Using the steady state density matrix Eq.(\ref{steady}) we can derive a formal expression for the 
DC Josephson current in the system:
\be
        I  = \frac{e}{T}  
         \mathbf{z} \cdot \left[ (\id -M_0)^{-1} \mathbf{v}_0 -
        \mathbf{z} \cdot (\id -M_{T/2})^{-1} \mathbf{v}_{T/2} \right]
        \label{formal_current}
\ee
where $\mathbf{z}$ is the unitary vector $(0,0,1)^{T}$, 
$\cdot$ stands for the usual scalar product in $\mathbb{R}^3$, 
and $M_{T/2}$, $\mathbf{v}_{T/2}$, 
$M_{0}$, $\mathbf{v}_{0}$ 
are defined in \eq{last_name}. 
Their explicit form is
\begin{widetext}
\bea 
        M_0 & = & \exp(G_{\law}t_{\law}) \cdot R(\chi_{\law}) \cdot
        \exp(G_R t_R) \cdot \exp(G_{\raw}t_{\raw}) \cdot R(\chi_{\raw}) 
        \cdot \exp(G_L t_L) \\
        \mbf{v}_0 & = & - 2 \exp(G_{\law}t_{\law}) \cdot R(\chi_{\law}) \cdot
        \left[ \gamma_R G_R^{-1} \cdot ( \id -\exp(G_R t_R)) \mbf{w}_R 
        + \gamma_L \exp(G_R t_R) \cdot \exp(G_{\raw} t_{\raw}) \right. \nonumber \\
        & & \left.  \cdot 
        R(\chi_{\raw}) \cdot G_L^{-1} \cdot ( \id -\exp(G_L t_L)) \mbf{w}_L \right] \, ,
\eea
\end{widetext}
and $M_{T/2}$, $\mathbf{v}_{T/2}$ are obtained from $M_{0}$, $\mathbf{v}_{0}$
by the exchange of right  and left Josephson contacts and of forward and backward 
free evolution time.  This means that  $M_{T/2}=\mathcal{P}M_{0}$, 
$\mathbf{v}_{T/2}=\mathcal{P}\mathbf{v}_{0}$ with $\mathcal{P}$ acting on the parameters $\theta_{L(R)},\chi_{\ra(\la)}, \gamma_{L(R)}t_{L(R)},\gamma_{\ra(\la)} t_{\ra(\la)}, \phi_{L(R)}$ as:
\be
        \mathcal{P}:\, (L,\ra,R,\la) \Rightarrow (R,\la,L,\la) \, .
        \label{transform}
\ee
The expression of the current, which depends on all the previous parameters
can be obtained analytically from Eq.(\ref{formal_current}) by explicitly 
writing $M_0$ and $\mathbf{v}_0$ 
in terms of the various parameters. The current depends only on the phase difference 
between the two superconductors $\phi_R-\phi_L$. It is an odd function with respect 
to the action of $\mathcal{P}$ defined by Eq.~(\ref{transform}).
From this observation follows that, even for $\phi=0$, there can be a supercurrent 
between the leads as long as  the evolution over a cycle is not $\mathcal{P}$-invariant.
In this sense the system behaves like a non-adiabatic Cooper pair pump.

\begin{figure*}[t]
\begin{center}
        \includegraphics[width=.45\textwidth]{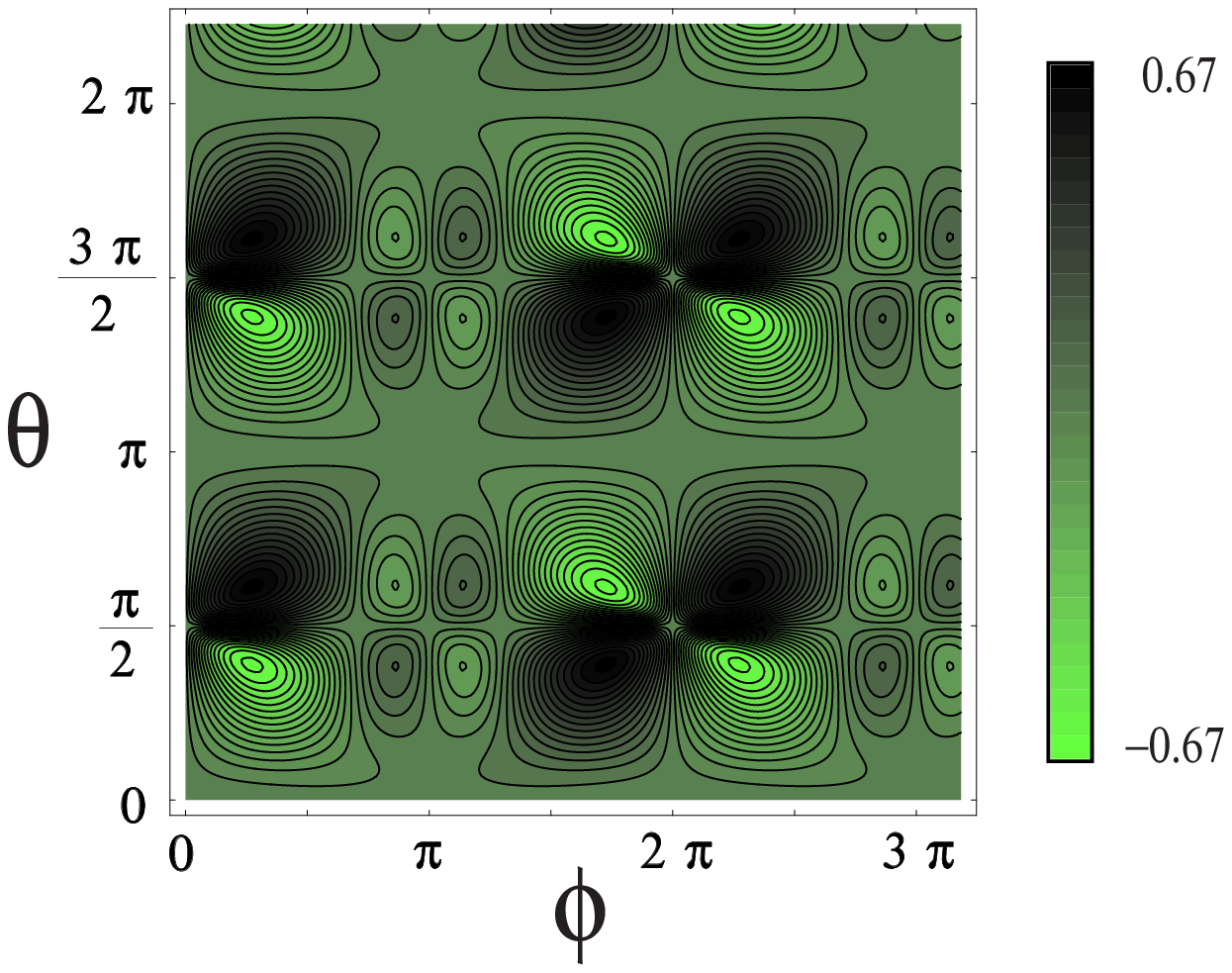}         \includegraphics[width=.45\textwidth]{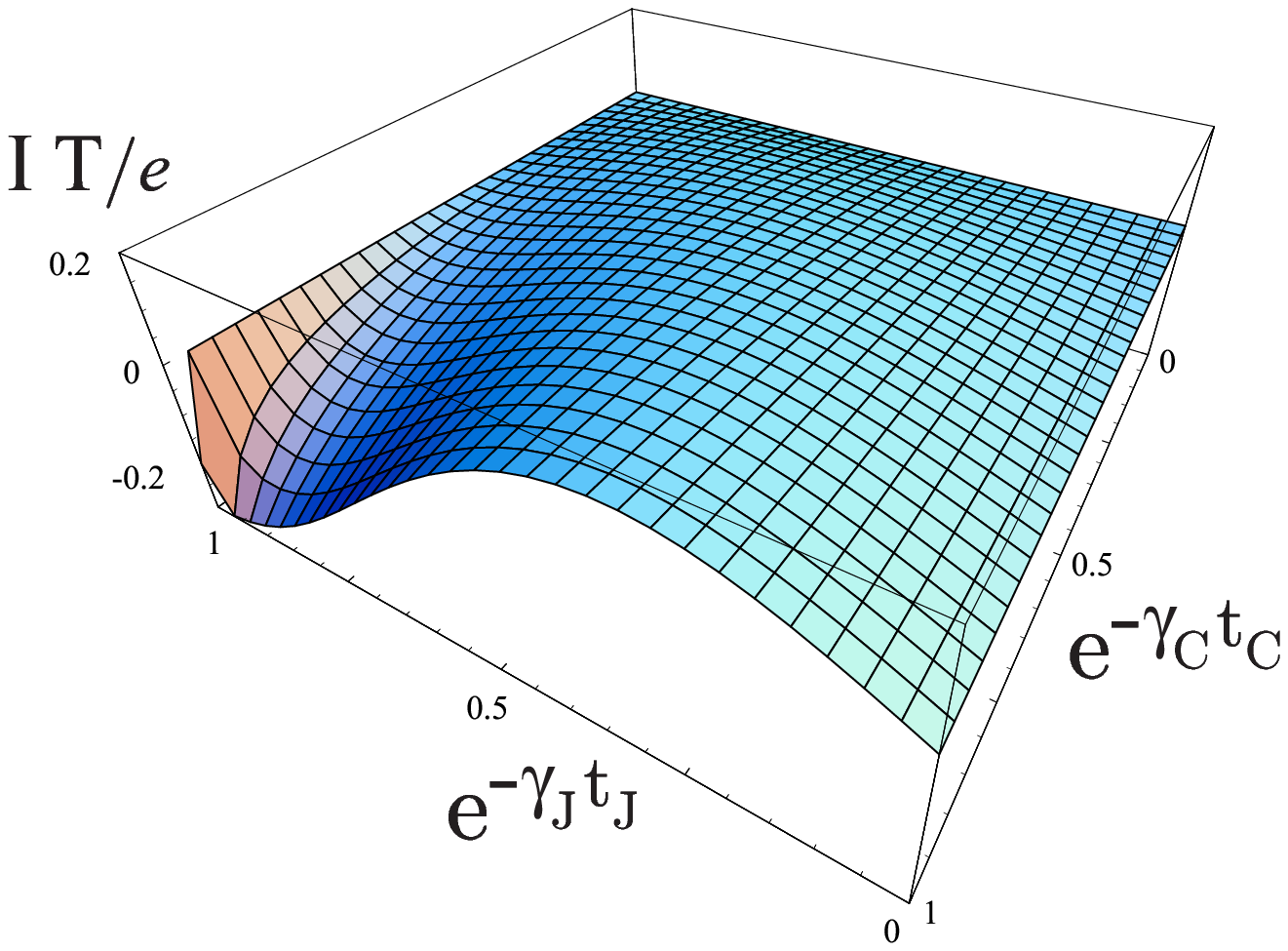}
        \end{center}
        \caption{(Color online). Left panel. Supercurrent (in units of $e/T$) as a function of the
        superconductor phase difference $\phi$ and of the phase
        accumulated during the contact to one of the electrodes $\theta$.
        The other parameters are fixed as: $\chi=5 \pi/6$, $e^{-\gamma_J
        t_J}=3/4$, $e^{-\gamma_C t_C}=4/5$. The plot is obtained for $T_b
        \ll E_J$. Right panel. 
        Average current ($T_b \ll E_J$) as a function of the dephasing
        rates, with $\phi=-3\pi /4$, $\theta=7 \pi /10$, $\chi=5 \pi/6$. As a
        function of $\gamma_J t_J$, the supercurrent has a not-monotonous
        behavior. Note the change of sign in the current obtained by
        varying decoherence rates in each time interval separately.} 
\label{corrente1}
\end{figure*}

The main features of the Josephson current in the Cooper pair shuttle have been 
discussed in Ref.~\onlinecite{romito03} and we recall them here 
for completeness however providing a number of  new results and additional details.
In the case of $\theta_L=\theta_R=\theta$, $\chi_{\ra}=\chi_{\la}=\chi$, 
$\gamma_L=\gamma_R=\gamma_J$, $\gamma_{\ra}=\gamma_{\la}=\gamma_C$, 
$t_L=t_R=t_J$, $t_{\ra}=t_{\la}=t_C$,
Fig.~\ref{corrente1} shows a typical plot of $I$ 
as a function of $\theta$ and $\phi$. 
Depending on the value of $\theta$ (a similar behavior is observed as a function of
$\chi$), the critical current can be negative, i.e. the system can
behave as a $\pi$-junction. The current dependence on the various phases is 
the result of the interference between different path corresponding 
to different time evolutions for the charge states in the grain.
By changing $\gamma_J t_J$ and $\gamma_C t_C$, certain interference paths are 
suppressed, resulting in a shift of the interference pattern and ultimately in a change of
the sign of the current, as shown in Fig.\ref{corrente1}.

Another interesting aspect of the Josephson current is that it is 
a non-monotonous function of $\gamma_J t_J$, {\it i.e.} by {\em
increasing} the damping, the Josephson current can {\em increase}.
The behavior as a function of the dephasing rates is
presented in  Fig.\ref{corrente1}. The presence of a maximum
Josephson current at a finite value of $\gamma_J t_J$ can be
understood by noting that the current is vanishing in 
the strong and weak damping limits.
In both the limits simple analytic expressions are available.


\begin{widetext}
If the dephasing is strong, $I$ can be expanded in powers of
$e^{-\gamma_{L(R)} t_{L(R)}}$ and $e^{-\gamma_{\ra(\la)} t_{\ra(\la)}}$ and, 
to leading order,
\begin{equation}
        I  \sim  \frac{2 e}{T} \left[ \tanh \left( \frac{E_L}{T_b} \right) 
        e^{-(\gamma_L t_L+\gamma_{\la} t_{\la})}
        \sin(2\theta_L) \sin(\phi-\chi_{\la}) +
        \tanh \left( \frac{E_R}{T_b} \right) 
        e^{-(\gamma_R t_R+\gamma_{\ra} t_{\ra})}
        \sin(2\theta_R) \sin(\phi+\chi_{\ra})\right] \; .
\label{general_strong} 
\end{equation}
For simplicity we assume that the Josephson energies at 
left and right contacts are equal, as well as the contact
and free evolution time. In this case the previous expression 
is simplified to
\be
I_{strong} \sim \frac{2 e}{T}
\tanh \left( \frac{E_J}{T_b} \right)
e^{-(\gamma_J t_J+\gamma_C t_C)} 
\cos(\chi_{\ra}+\chi_{\la})\sin(2\theta)  \, 
\sin(\phi+(\chi_{\ra}-\chi_{\la})) \; . 
\label{not_so_general_strong}
\ee
\end{widetext}
It is worth to notice the presence of a net DC current even
in the case of $\phi=0$ as argued from general argument presented before. 
The role of breaking the $\mathcal{P}$-invariance
is then played by the difference of the dynamical 
phases accumulated in the forward and backward 
free evolution time intervals, $\chi_{\ra}-\chi_{\la}$.
If instead we assume a perfect $\mathcal{P}$-invariance, 
we recover the known expression for the 
DC current of Ref.~\onlinecite{romito03}
\begin{eqnarray}
        I_{strong} & \sim & \frac{2 e}{T}
        \tanh \left( \frac{E_J}{T_b} \right)
        e^{-(\gamma_J t_J + \gamma_C t_C)} \nonumber \\
        & & \cos(2\chi)\sin(2\theta)  \, \sin\phi \, .
\label{limit0}
\end{eqnarray}
Strong dephasing is reflected in the simple ({\it i.e.} $\propto \sin
\phi$) current-phase relationship and in the exponential
suppression of the current itself. Strong dephasing, in fact, suppresses 
coherent transport over multiple cycles, thus giving a corresponding suppression 
of higher harmonics in the current-phase relationship,
{\it i.e.} a suppression of terms $\propto \sin^{2m+1}(\phi)$, $m \in N$. \\
For the sake of simplicity, from now on, we present all the result 
in the case of perfect $\mathcal{P}$-invariance of the time evolution 
of the density matrix in a period. 
This is not a serious limitation for the experimental setups.

In the opposite limit of weak damping defined by $\gamma_J t_J \ll
\gamma_C t_C \ll 1 $~
\begin{equation}
        I_{weak} \sim
        \frac{2e}{T}
        \tanh \left(\frac{E_J}{T_b} \right)
        \frac{\gamma_J t_J}{\gamma_C t_C}
        \frac{(\cos\phi+\cos 2\chi)\tan\theta \sin\phi}
        {1+\cos\phi \cos2\chi} \, .
\label{limit1}
\end{equation}
The current tends to zero if the coupling with the bath is
negligible during the contact time. In this case the time evolution in the 
intervals $L,R$ is almost unitary, while, in the region $C$, pure dephasing 
leads to a suppression of the off-diagonal terms of the reduced density matrix 
$\rho (t)$. As a result, in the stationary limit the system is described by a
complete mixture with equal weights. At the point $(\gamma_J t_J, \gamma_C t_C) = (0,0)$
our model is not defined as discussed at the end of Section~\ref{calcoli}.
The limiting value of the current in approaching such point  depends on the 
relative strength  $\gamma_J t_J  \lessgtr \gamma_C t_C$ 
between this two parameters.

The current tends to zero in both limiting cases of large and
small $\gamma_J t_J$. Therefore one should expect an optimal coupling
to the environment where the Josephson current is maximum.
A regime where the crossover between the strong and weak damping
cases can be described in simple terms is the limit $\gamma_C
\rightarrow 0$, for a fixed value of $\theta$. For example, at
$\theta=\pi /4$ the current reads
\begin{widetext}
\begin{equation}
        I = \frac{2e}{T}
        \tanh \left(\frac{E_J}{T_b} \right)
        \frac{2e^{-\gamma_J t_J}[2e^{-2\gamma_J t_J}\cos\phi +
        (1+e^{-4\gamma_J t_J})\cos2\chi]\sin\phi}
        {(1+e^{-2\gamma_J t_J})(1+ e^{-2\gamma_J t_J}\cos\phi \cos2\chi
        +e^{-4\gamma_J t_J})}
\label{crossover}
\end{equation}
\end{widetext}
In the limit of vanishing $\gamma_J t_J$, Eq.(\ref{crossover})
corresponds to the situation discussed in Ref.~\onlinecite{gorelik01}.
Indeed, both expressions are independent of the dephasing rates.
The difference in the details of the current-phase(s) relationship
are due to the different environment.

In all the three cases presented here,
Eqs.(\ref{limit0}, \ref{limit1}, \ref{crossover}), the change of
sign of the current as a function of the phase shifts $\theta$ or
$\chi$ is present.

\subsection{Effect of driving fluctuations}
\label{motion_fluctuation}

The expressions for the current, discussed in the previous Section, depend 
on the specific form of the coupling between the system and the reservoir. 
The decoherence we considered so far originates from gate voltage fluctuations.
In addition in the shuttling mechanism an unavoidable coupling to an environment
producing fluctuations in the period and shape of the driving is also present. 
We are therefore interested also in considering the effect of fluctuations in the 
time dependence of the external parameters on the Josephson current.
Having assumed a step-like time dependence of the parameters of the Hamiltonian, 
noise in the external driving consists in  fluctuations of the switching times. 
This means that the contact times $t_L$, $t_R$ and the free evolution times 
$t_{\ra}$, $t_{\la}$ take, at any cycle $i$, a random value $t_b(i)$, $b=L, R ,\ra, \la$. 
It is reasonable to assume that the fluctuations of any switching time are independent 
on the others. In terms of time intervals $t_b(i)$, it follows that 
the fluctuations  $\Delta t_b(i)$ around the average value $t_b$, are uncorrelated 
at different periods. Within the same period the fluctuations of any two distinct time 
intervals are also independent. Hence 
\be
        \motavg{\Delta t_{b}(i)}=0 
\ee
and 
\be
        \motavg{(\Delta t_a(i))^n (\Delta t_b(j))^m}=
        \motavg{(\Delta t_a(i))^n} \motavg{(\Delta t_b(j))^m}        
\label{motion_distribution}
\ee
if $ i \neq j \vee a \neq b$. The integer valued arguments of $\Delta t_b(\cdot)$ 
labels the periods and the subscript index runs over the 
set ${L,R, \ra, \la}$.  The average on the 
stochastic process is defined by $\motavg{\phantom{\cdot}}$. We will discuss later the 
distribution function of $\Delta t_b (i)$.

By using the same notation in \eq{map}, we can write the evolution of the density 
matrix after a finite  number of cycles, $h \geqslant 1$, as
\begin{widetext}
\be
        \mbf{r}(t+\sum_{k=1}^h T(k)) =  
        \prod_{\lambda=0}^{h-1} M_t(\lambda) \mbf{r}(t) + 
        \sum_{\lambda=0}^{h-2} \prod_{\mu=\lambda}^{h-2} M_t (\lambda+\mu+1) 
        \mbf{v}_t(\lambda) +
        \mbf{v}_t (h-1) \, .
\label{complicated_evolution}
\ee 
\end{widetext}
In the previous equation the expressions $M_t (i)$ and $\mbf{v}_t (i)$ 
are those defined in \eq{map}. The index in parenthesis indicates the explicit 
dependence of both $M_t$ and $\mbf{v}_t$ on various $t_b(i)$.
We refer to $T(k)=\sum_bt_b(k)$ as the period although the time evolution is 
no longer periodic (before averaging); $T(k)$ is in fact the time the 
shuttle takes to complete a cycle, $k$ labels the number of cycles. 
The first term in the right-hand-side of \eq{complicated_evolution}
vanishes in the long-time limit $h \raw \infty$.
Averaging the previous expression over the fluctuations of the switching times according 
to \eq{motion_distribution} is straightforward, leading to
\be
        \motavg{\mbf{r}_{\infty}(s)} = \left( \id - \motavg{M_s} \right)^{-1}
        \motavg{\mbf{v}_s} \, .
\label{driv_fluct_vect}
\ee
Note that if we had considered only the external driving fluctuations ({\it i.e.} 
neglecting the effect of gate voltage fluctuations) we would have found, in the 
steady state,  $\rho_{\infty}(t) \propto \id$. 
In this case in fact the evolution of $\mathbf{r}(t)$ consists in an alternate sequence 
of rotation on the Bloch sphere around the $(1,0,0)$ and $(0,0,1)$ axes.
Due to uncertainty in the rotation angles
it is a random walk which leads, at long time, to a uniform distribution 
over the Bloch sphere. The nontrivial result in \eq{driv_fluct_vect} arises 
because of the interplay between the two stochastic 
processes of gate voltage fluctuations and switching time fluctuations.
They are independent because there is a time scale separation 
between these two processes: Correlations in the quantum bath do occur on a 
time scale $\tau_c \ll T$, while the time intervals $t_b(i)$ do not fluctuate 
within any single cycle. 

We are interested in averaging the current after 
the system has reached its steady state 
\begin{widetext}
\be
        \motavg{I} = \lim_{N \ra \infty}  \left\langle 
        \frac{2e}{\sum_{i=1}^N T(i)}  \mbox{Tr}  \left[ \hat{n}
        \sum_{j=1}^N \left( \mathcal{M}_{0 \ra \sum_{h=1}^{j-1} T(h)+t_L(j)} 
        -  \mathcal{M}_{0 \ra \sum_{h=1}^{j-1} T(h)} \right)[\rho(0)] 
        \!\!\! \!\!\!\! \phantom{\sum_{j=1}^N}
        \right] \right\rangle_t \, ,
\label{current_formal_fluct}
\ee
\end{widetext}
In order to proceed further we need to specify the distribution function
$P(\Delta t_b(i))$. Let us note that the
distribution is meaningful only if $P(\Delta t_b(i))=0$ for $\Delta t_b(i)<0$.
We consider 
\be
        P(\Delta t_b(i))=\theta(\Delta t_b(i) + \tau)
        \theta (\tau- \Delta t_b(i))/2\tau\, ,  
\ee
with $\tau < t_b \, \forall b$ as a toy model: the underlying physical
idea is that the switching time can be controlled with a 
precision $2\tau$ and the switching can happen with equal probability in the interval 
$[t_b-\tau, t_b+\tau]$\footnote{A Gaussian distribution 
function for the variables $\Delta t_b(i)$ cannot be assumed;
it is inconsistent with the previous condition, 
$\mathcal{P}(\Delta t_b(i))=0$ for $\Delta t_b(i)<0$, and, in fact,
 gives rise to incurable divergences due to the presence
of the term $1/T$ in the expressions to be averaged.}. 
We do not expect that this simple form of the distribution function can determine 
the quantitative details of the Josephson current, rather it can grasp the main 
features of physical effects due to imprecision in controlling the 
external driving. For the sake of concreteness, let us consider the 
limit of strong dephasing (\eq{limit0}), when the expression for the current 
considerably simplifies. The strong dephasing leads to a rapid 
loss of memory of the initial conditions. One can suppose that this occurs after 
one cycle independently on the averaging process. It follows that, in \eq{current_formal_fluct}, 
\begin{widetext}
\be
\motavg{I} = \lim_{N \ra \infty}  \left\langle 
        \frac{2e}{\sum_{i=1}^N T(i)}  \mbox{Tr}  \left[ \hat{n}
        \sum_{j=1}^N \left( \mathcal{M}_{ \sum_{h=1}^{j-1} T(h)  \ra \sum_{h=1}^{j-1} T(h)+t_L(j)} 
        -  \id \right)[\rho_{\infty}(0)] 
        \!\!\! \!\!\!\! \phantom{\sum_{j=1}^N}
        \right] \right\rangle_t \, ,
\ee
$\mathcal{M}$ depends only on the  stochastic parameters  of the last ($=j$th) cycle 
and $\rho_{\infty}(0)$ on parameters of the $j-1$th cycle.
By considering the expansion of  the denominator in the previous equation as 
\be
        1/(\sum_{i=1}^N T(i)) = 
        1/(NT)(1-\sum_{i=1}^N\sum_b \Delta t_b(i)/NT+ \dots)
\ee
together with \eq{motion_distribution}, \eq{current_formal_fluct}
reduces to
\be
\motavg{I} = \lim_{N \ra \infty}  \frac{2e}{NT}\left\langle 
          \mbox{Tr}  \left[ \hat{n}
        \sum_{j=1}^N \left( \mathcal{M}_{ \sum_{h=1}^{j-1} T(h)  \ra \sum_{h=1}^{j-1} T(h)+t_L(j)} 
        -  \id \right)[\rho_{\infty}(0)] \!\!\! \!\!\!\! \phantom{\sum_{j=1}^N}
        \right] \right\rangle_t + \mathcal{O}\left( 1/N \right) \, .
\ee
In the $N \raw \infty$ limit, only the first term in the 
previous equation is non-vanishing. 
It means that, as a consequence of the strong dephasing, 
the effect of fluctuations in the term $1/T$ in the definition of the current 
are ineffective. The final expression for the average current is
\be
        \motavg{I} = \frac{2e}{T} \sin\phi 
        \motavg{e^{-\gamma_C t_{\la}} \cos \frac{E_C t_{\la}}{\hbar}}
        \motavg{e^{-\gamma_J t_L} \sin \frac{E_J t_L}{\hbar}}
\ee
which, in the lowest orders in the small parameter $\tau E_C/\hbar$, reads
\be
        \motavg{I}  \approx \frac{2e}{T}\left\{ \left[ 
        1-\frac{1}{6}  \, \left( \frac{E_C \tau}{\hbar}\right)^2 \right] 
        I_{strong} + \frac{1}{3}  \, \left( \frac{E_C \tau}{\hbar}\right)^2 \left[  
        \frac{\hbar \gamma_C}{E_C} 
        \times \sin(2 \theta) \sin(2 \chi) 
        - \, \frac{E_J}{E_C} \,\frac{\hbar \gamma_J}{E_C} \, \cos(2 \theta) 
        \cos(2 \chi) \right] \sin \phi \right\}\, , 
\label{final_motion_fluct}
\ee
\end{widetext}
with the further condition 
$ 1/T \ll \gamma_J \ll E_J/\hbar \ll  E_C/\hbar \gg \gamma_C \gg 1/T$.
The current $I_{strong}$ is defined as the current in absence of driving fluctuations 
and is that obtained in \eq{limit0}.
The leading order correction to $I_{strong}$ does not modify the functional dependence 
of the current on the dynamical phases: It is a simple renormalization
of the pre-factor, $1 \ra 1-(E_C \tau)^2/(6 \hbar^2)$.
\eq{final_motion_fluct} shows that higher order corrections can instead lead to 
a modifications of the functional dependence of the current on dynamical phases.

\subsection{AC Josephson effect}
\label{ac}
When a bias voltage is applied to a Josephson junction it results in 
an alternating current. This is the AC Josephson effect which can 
be derived from the expression for the Josephson current $I=\frac{2e}{\hbar} E_J 
\sin(\phi_R-\phi_L)$ supplemented by the Josephson relation 
$\frac{d}{dt}(\phi_R-\phi_L)=\frac{2e}{\hbar}V$. In the present problem 
the alternating current will not be simply sinusoidal, it is therefore convenient 
to consider its frequency spectrum defined as
\be
\tilde{I}(\omega)=\int_{\mathbb{R}} dt \, e^{-i\omega t} \avg{I}(t) \, .
\ee
(in the simple case of a sinusoidal current it reads
$\tilde{I}(\omega) \propto \delta(\omega -2 e V/\hbar)$ for $\omega > 0$).

The application of a finite voltage bias in the Cooper pair shuttle gives rise 
to a quite rich situation, due to the interplay of of the voltage bias effect 
and the underlying periodic motion of the shuttle. The relative magnitude of the 
characteristic frequencies, $\sim 2 e V/\hbar$ and $\sim 1/T$, determine 
different regimes that we are going to investigate. 

We  set the electric voltage of the left and right electrode respectively equal 
to $V_L=-V/2$ and $V_R=V/2$. There is no dissipative current through the system 
as long as $eV \ll 2 \Delta$. In the limit  $C_L,C_R \lesssim C_g$, and 
$(V/2) \ll V_g$ the Hamiltonian of the system is still given by 
\eq{h0}. The effect of the electric potential $V_{L(R)}$ in the two leads can be included,
by means of a gauge transformation, into time dependent 
phases of the condensate wave functions, 
$\ket{\psi_{L(R)}} \raw e^{2 i e V_{L(R)} t/\hbar} 
\ket{\psi_{L(R)}}$, or equivalently, 
$$
        \phi_{L(R)} \raw \phi_{L(R)}+ 2 e V_{L(R)} t/\hbar \,\, .
$$ 
The Hamiltonian  describing the effect of voltage 
bias in the Cooper pair shuttle becomes 
\bea
        \hat{H}_{AC} &=& E_C(t) [\hat{n}- n_g(t)]^2 
        \nonumber \\
        &-& E_J \Theta_L(t)\cos(\hat{\varphi}-\phi_L-eVt/\hbar)
        \nonumber \\  
        &-& E_J \Theta_R(t) \cos(\hat{\varphi}-\phi_R+eVt/\hbar)\, .
\label{AC_hamiltonian}
\eea
We are interested in the frequency dependent current
\bea
        \tilde{I}(\omega) &=& 2 e \frac{E_J}{\hbar} \int_{\mathbb{R}} dt \, 
        e^{-i \omega t} \Theta_L (t) \mathcal{I}(t) 
        \nonumber \\
        &=& 2 e \frac{E_J}{\hbar} \sum_{k \in \mathbb{Z}} e^{-i k \omega T} 
        \int_0^{t_L} ds \, e^{-i \omega s} \mathcal{I}(s+kT) \, .
\label{frequency_current}
\eea
 where $\mathcal{I}(t)=\mbox{Tr} \left\{ \sin \left( \hat{\varphi} 
-\phi_L  -e V t/\hbar \right) \rho(t) \right\}$, 
according to the definition of current in Eq.~(\ref{current_L}).

\begin{figure}[t]
\begin{center}
\includegraphics[width=0.45\textwidth]{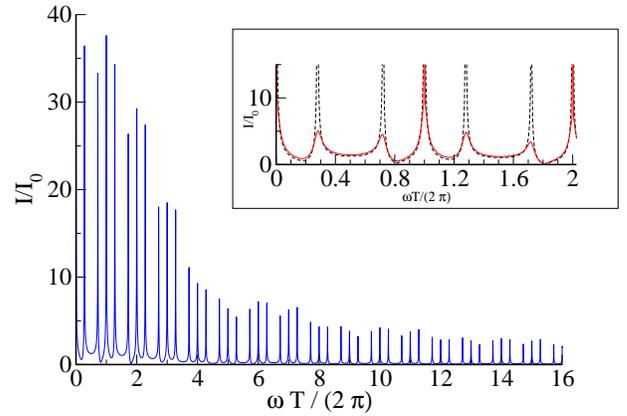}
\end{center}
\caption{(Color online). Absolute value of the Fourier transform
 of the Josephson current in units of 
 $I_0=2eE_J/\hbar$. The plot is obtained 
for $t_J=t_C=T/4$, $E_C T/\hbar=60$, 
$E_J T/\hbar=6$, $\gamma_J T=\gamma_C T= 0.001$.
In the inset we plot the Fourier transform of the Josephson 
current in a restricted range of frequencies for 
$\gamma_J T=\gamma_C T= 0.001$ (black dotted line) and 
for $\gamma_J T=\gamma_C T= 0.1$ (full red line).}
\label{frequency_current_1}
\end{figure}

We computed numerically the time evolution of the density matrix 
of the island and obtained from it the frequency spectrum of 
the Josephson current. As a warm up we consider the
simplest case which consists in neglecting the effect of voltage bias. 
This is the same case considered in previous Sections,
in which, however, we analyze the instantaneous current
rather then the averaged one. 
The results are presented in Fig~\ref{frequency_current_1}.
The spectrum clearly signals the periodicity of the time dependence of 
the current signal and of the presence of the  $\Theta_L$ function.
In the steady state the current is repeats periodically and then 
$\mathcal{I}(s+kT)=\mathcal{I}(s)$ in Eq.~(\ref{frequency_current}).
As a consequence the current spectrum presents peaks at integer multiples 
of the frequency $\omega_n=2 \pi/T$, 
\be
        \tilde{I}(\omega) = \sum_n  \tilde{A}_n
        \delta \left( \omega-2\pi n/T \right)
\ee
with
$$
        \tilde{A}_n=\int_0^{t_J} ds \, 
        \exp(-2 \pi i n s/T) \, \mathcal{I}(s) \,\,\, .
$$
The form of $\tilde{A}_n$ is fixed by the expression of the density matrix at the 
fixed point through $\mathcal{I}(s)$, $\mathcal{I}(s)= \mbf{y} \cdot R_z(\phi_L) \mbf{r}(s)$,
and with $\mbf{r}(s)$ determined by \eq{solution}.
The signal $\mathcal{I}(s)$ consists of damped oscillations at frequency $E_J/\hbar$,
$\mathcal{I}(s) \propto \exp(-\gamma_J s)  \sin(E_J s/\hbar+\alpha_0)$. 
The Fourier transform of such a signal determines the characteristic features of the 
current spectrum presented in \fig{frequency_current_1}. Now
$|\tilde{A}_n|$ displays oscillations at frequencies  $T/t_J$ modulated by a power-law 
decay function ($\propto 1/\omega$, for $\omega \gg E_J/\hbar $). 
The nature of the other peaks shown in the inset in \fig{frequency_current_1} is 
completely different, in fact they are strongly suppressed 
by the presence of decoherence. We can describe the mechanism that
originates these peaks if the decoherence
is absent. In this case any component of the density matrix
is an oscillatory function with frequency $E_J/\hbar$, $E_C/\hbar$ at contact 
or free evolution interval respectively. The matching between these two 
different functions at the boundary between consecutive time intervals determines the phase 
$\alpha_k$ in the expression for $\mathcal{I}(s)$ after $k$ periods. We get
$\mathcal{I}(s) \propto \sin(E_J s/\hbar + k T (E_C-E_J)/(2\hbar))$. Substituting 
this expression into \eq{frequency_current}, we find the existence of peaks in the 
frequency spectrum of the current at $\omega = (2 \pi/T) (k \pm (E_C-E_J)/4 \pi \hbar)$, 
$k \in \mbb{Z}$, as presented in \fig{frequency_current_1}.
The stronger decoherence is, the more suppressed such peaks are.

\begin{figure}[t]
\begin{center}
\includegraphics[width=0.45\textwidth]{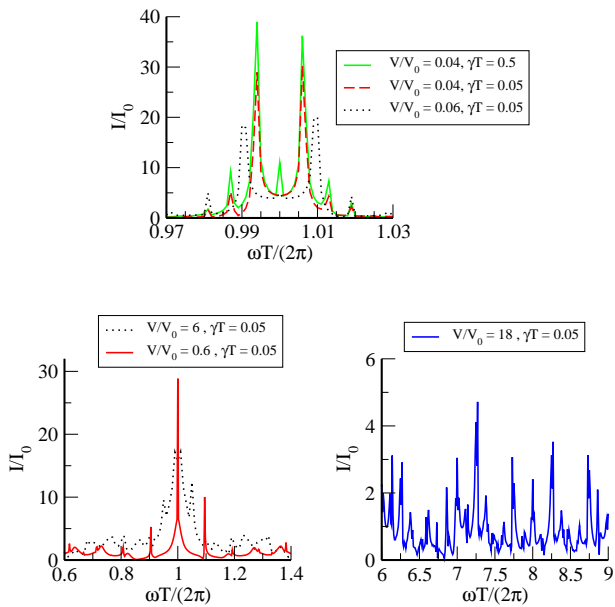}
\end{center}
\caption{(Color online). Absolute value of the Fourier transform
        of the Josephson current (all panels) in units of 
        $I_0=2eE_J/\hbar$ 
        in the different regimes: $eVT/\hbar \ll 1$ (upper panel),
        $eVT/\hbar \simeq 1$ (lower left panel), 
         $eVT/\hbar \gg 1$ (lower right panel).
        The plots are obtained 
        for $t_J=t_C=T/4$, $E_C T/\hbar=60$, 
        $E_J T/\hbar=6$.
        The values of $V$, in units of $V_0=\hbar/(2eT)$, and  
        $\gamma_J=\gamma_C=\gamma$ for the various plots
        are indicated in the legends.}
\label{voltaggio}
\end{figure}

This picture is modified in presence of a 
finite voltage bias between the two superconductors, as 
presented in \fig{voltaggio}.
The features of the  frequency spectrum in this case depends on whether the 
condition $2eV \ll \hbar/T$ or $2eV \gg \hbar/T$ is fulfilled. 
For small $V$, one can suppose that the system reaches a steady state  which evolves in 
time only through the time dependence of the parameter $\phi \ra \phi+2e VT/\hbar$. 
It means we can replace $\mathcal{I}(s+kT) \sim \sin (E_J s /\hbar + 2e V k T/\hbar)$. 
In this regime the spectrum  exhibits a splitting
of the frequencies of the peaks $\omega = (2\pi/T)( k \pm eVT/(\pi\hbar))$.
This is shown in the upper panel in \fig{voltaggio} for $k=1$, the peak at lowest frequency. 
The effect  the peaks at higher frequencies is the same, therefore we focus on the first 
peak at $\omega = 2\pi/T$ since it is the most pronounced 
(the values at the  peaks $\tilde{A}_n$ decays with increasing the frequency
as in the case $V=0$, see \fig{frequency_current_1}). 
\fig{voltaggio} also shows the presence of other peaks at integer multiples of $2eV/\hbar$
which cannot be interpreted within the  simple model just presented.
Moreover the width of the peaks is weakly dependent on $\gamma_{J(C)}$.
The presence of higher harmonics in the spectrum becomes evident for $\hbar/2eV \sim T$ 
(see lower left panel in \fig{voltaggio}). 
The presence of higher harmonics, although we do not have a detailed simple explanation, 
is expected once we write the Josephson coupling in the Hamiltonian \eq{AC_hamiltonian} as
$E_J \cos (\phi-2e V/\hbar) \cos(\hat{\varphi}+(\phi_L+\phi_R)/2))$. 
This shows that $E_J$ is modulated by an oscillatory time dependent factor.
Even if further complicated by the presence of time order operator in the time evolution
operator, the current $\mathcal{I}(s+kT)$ would present terms $\propto \sin(E_J 
\cos(2e V(s+ Kt)/\hbar))$ which generates all the harmonics. In the voltage dominating 
regime $2eV/\hbar \gg 1/T$, the spectrum does not display new effects apart 
from the described $V$-splitting and  an enhancement of peaks corresponding to 
high harmonics (lower right panel in \fig{voltaggio}).

\section{The chaotic regime of Cooper pair shuttling}
\label{chaotic}

Up to now we have discussed the properties of the Cooper pair shuttle in the 
limit of small Josephson coupling, $E_J \ll E_C$. In this Section we consider 
the opposite regime, $E_J \gg E_C$. If the external time-dependent driving 
were absent, such limit would not  be of great interest: It simply corresponds 
to a SSET in the Josephson dominate regime, whose physics is already known~\cite{tinkham96}.
Due to the time dependence of the Josephson couplings in the Cooper pair shuttle, 
instead, there is a time lapse in which the Josephson energy is vanishing 
(see \fig{evolution}), and therefore the charging effects still play a leading role.

We will show that the dynamics of the Cooper pair shuttle mimics that of a Quantum 
Kicked Rotator (QKR) with the additional free parameter, $\phi$. The kicked rotator is 
a chaotic system in the classical limit. In the quantum regime it presents a variety 
of interesting phenomena, including dynamical localization~\cite{izrailev90}. 
Despite the long-standing interest in the QKR, only few proposals have been put 
forward and the only experimental implementation so far has been realized with  cold 
atoms exposed to time-dependent standing waves of light~\cite{moore95, ammann98, zhang04}. 
The Cooper pair shuttle can therefore provide a remarkable implementation of the 
QKR~\cite{romito05'}  by means of superconducting nanocircuits\footnote{Periodically 
driven Josephson junction have been already suggested to study quantum chaos 
in Ref.~\onlinecite{graham91}.}. This allows us to study the effects of dynamical localization on the transport properties of a mesoscopic device. So far the effect of dynamical localization on the current in a mesoscopic systems has been discussed only in Ref.~\onlinecite{basko04}  for a quantum dot under AC pumping in which it affects the shape of Coulomb blockade peaks.

\subsection{From classical to quantum dynamics in the chaotic Cooper pair shuttle}
\label{JQKR_dynamics}

We start our analysis from the Hamiltonian in \eq{h0} which is
valid irrespective of the relative strength between $E_C$ and $E_J$.
For sake of simplicity we assume $n_g=0$ throughout 
this section , thus rewriting \eq{h0}
as 
\bea
\hat{H}_0 & = & E_C \hat{n}^2 - E_J \sum_{n \in \mathbb{N}}
\left[  \cos (\varphi-\phi_L) \Theta (t-n T) + \right. \nonumber \\
  & + &\left. \cos (\varphi -\phi_R) \Theta (t- (2n+1) T/2)  \right] \, ,
\label{kicked_rotator}
\eea
where $\Theta (t)=\theta(t)\theta(t_J-t)$ having assumed  the simple limit 
$t_L=t_R=t_J$, $t_{\ra}=t_{\la}=t_C$. The dynamics of the Cooper pair shuttle reduces to 
that of a QKR when the contacts times are short enough  to neglect the effects 
of charging energy. In this case the time evolution operator during the Josephson 
contact is $\exp[-\imath E_J t_J\cos(\hat{\varphi}-\phi_{L,R})/\hbar]$. In the rest of 
the cycle the shuttle evolves according to the charging Hamiltonian only.
Let us explain this in some detail. Because we are interested in the dynamics of the system 
at long time compared with the period $T$, any physical observable, 
as already noted in Section~\ref{calcoli}, is determined by the density matrix 
at integer multiples of the period and therefore by the Floquet operator which is 
the time evolution operator over a period. Under the assumption that $E_J \gg E_C$, 
if $E_C$ cannot induce a significant change of ${\varphi}$ during the Josephson contact, 
the Floquet operator becomes 
\be
        \hat{U}(T,0)=\hat{J}_R\hat{V}\hat{J}_L\hat{V} 
\label{4evolution}
\ee
where $\hat{J}$ and $\hat{V}$ are the time evolution operators
respectively during the Josephson contacts and the free evolution 
times
\be
        \hat{J}_{R,L}=e^{-\imath k\cos(\hat{\varphi}-\phi_{L,R})} \, , 
        \hat{V}=e^{-i\frac{K}{2k}\hat{n}^2} \,\, ,
\ee
where
$$
k=E_J t_J/\hbar \;\;\;\;\; \mbox{and} \;\;\;\;\; K=(2 E_C t_C/\hbar)(E_J t_J/\hbar) \, .
$$
The condition that the superconducting phase  difference is left unchanged at the contacts 
reads $n E_C t_J/\hbar \lesssim 1$~\cite{klappauf99}, thus establishing a condition  
on the maximum number of allowed charge states involved in the dynamics. This condition 
is essentially independent on the exact time dependence of $E_J(t)$.
The Floquet operator in \eq{4evolution} is the same of the Quantum Kicked Rotator (QKR)
with the additional parameter $\phi=\phi_R -\phi_L$. This shows that the physics
of the Cooper pair shuttle, in the limit $E_J \gg E_C$, may reproduces that of 
the QKR, and, for $\phi \neq 0$, provides an interesting generalization.

The kicked rotator is the first model in which characteristic quantum effects of 
classically chaotic system have been observed numerically~\cite{izrailev90,casati79}.
As the parameters $k$, $K$ in \eq{4evolution} are varied, the dynamics of the QKR 
exhibits several appealing phenomena, including quantum ergodicity, quantum resonances 
and dynamical localization~\cite{izrailev90}.
For a  discussion about classically chaotic system and various characteristic
features of their quantum version we refer to the literature (see Ref.~\onlinecite{izrailev90}. 
and references therein). Here we note that the exponential localization 
of the wave function due to interference effects is one of the most relevant of 
the mentioned features.  The dynamics of the quantum kicked rotator follows the 
classical exponential instability (characterized by a positive
Lyapunov exponent $\lambda$) up to the Ehrenfest time $t_E$~\cite{berman78}.
This sets the time scale at which quantum interference effects starts to be important 
leading to weak localization correction to the classical behavior~\cite{tian04}.
After a localization time $t^\star$ the classical-like diffusive
behavior is suppressed by quantum effects~\cite{izrailev90,casati79}. 
Since typically  $t^\star\gg t_E$, the diffusive behavior is  possible also in the 
absence of exponential instability.

The Floquet operator in \eq{4evolution} can be written 
through the redefinition $\hat{p}= (K/k) \hat{n}$ in terms of
\bea 
&& \hat{V}'=\exp(-i \hat{p}^2/2\kbar) \\
&& \hat{J}_{L(R)}'=\exp[-iK\cos(\varphi-\phi_{L(R)})/\kbar] \, ,
\eea
so that $\hat{U}(T,0)$ depends
only on $K$, while $[\hat{p},\hat{\varphi}]=-i K/k=-i \kbar$;
$\kbar$ plays the role of an effective Plank constant.
The classical limit is therefore obtained for $k\to \infty$ 
(or equivalently $\kbar \ra 0$), keeping 
$K={\rm const}$ and the classical dynamics 
depends only on the parameters $K$ and $\phi= \phi_R -\phi_L$. 

In the classical limit the equations of motion of the Cooper pair shuttle 
correspond to a slightly modified Chirikov map 
\be
        \left\{ \begin{array}{ccl}
        p_t & = & p_{t-1}- K \sin\left\{\theta_{t-1} - 
        [(t+1)\mbox{~\small mod~} 2]\phi \right\}\\
        \theta_t & = & \theta_{t-1}+ p_t, 
        \end{array} \right. \, ,
\label{classical_map}
\ee 
with $\theta=\varphi-\phi_L$ and where subscripts label the number of 
contacts with the leads (kicks).
The dynamics of a given distribution function in the phase space under the 
action of the Chirikov map at $\phi=0$ is known: 
For $K>1$, the angular variable $\theta$ is uniformly
spread over $[0,2\pi]$ after few kicks; $p$ follows a diffusive behavior.
The role of $\phi$ in the classical map can be  investigated by following 
Ref.~\onlinecite{rechester81}.  The idea is to substitute the deterministic 
description in Eq.~(\ref{classical_map}) with a probabilistic one where 
a random term $\delta \theta_t$ is added to the second equation
of (\ref{classical_map}). We obtain a diffusive dynamics 
for the charge on the central  island for $K>1$  (details of the calculation are in 
Appendix~\ref{diffusion_corrections}): 
$\avg{(n_{t}-n_0)^2} \stackrel{t\raw \infty}{\longrightarrow} 
 2 D t$, with $t$ measured in integers multiples of the period 
 and where $D$ is the diffusion coefficient 
\bea
        D &=& \frac{k^2}{2}\left[ 1-2 \cos (2 \phi) 
        J_2(K) + \mathcal{O}\left(\frac{1}{K} \right) \right] \, .
\label{phase_effect}
\eea 

\begin{figure}[t!]
\begin{center}
        \includegraphics[width=0.45\textwidth]{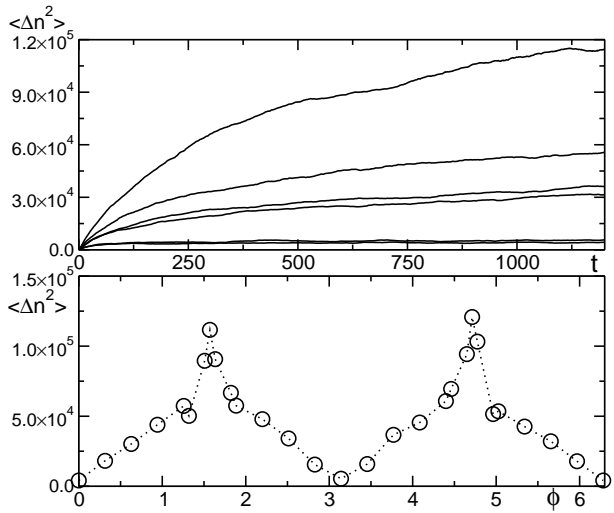}
        \caption{Upper panel. 
        $\langle (\Delta n)^2\rangle =\langle (n-\langle n\rangle)^2\rangle$ 
        as a function of time for $K=10$, $k=15$, 
        and phase difference (from bottom to top) 
        $\phi/(2\pi) = 0, 0.05,
        0.1, 0.4,$ $ 0.8, 0.25$. 
        Lower panel. Saturation value of 
        $\langle(\Delta n)^2\rangle\propto \ell^2$ 
        as a function of $\phi$ for the same parameter values 
        as in the upper panel.} 
\end{center}
\label{localizationfig}
\end{figure}

The QKR follows the classical diffusive behavior up to the localization
time $t^\star$. Quantum interference effects, as shown in Fig.~\ref{localizationfig}
(upper panel), for $t>t^\star$, suppress this chaotic diffusion: The wave function is 
exponentially localized in the charge basis, over a localization length $\ell$, and we 
have $\ell\sim t^\star \sim D$~\cite{izrailev90}.

The charge fluctuations of the central island freeze in time. The localization 
length can be further tuned by changing the phase difference as 
demonstrated in the lower panel of Fig.~\ref{localizationfig}.
Such quantum effects  have not yet found a complete analytic explanation.
An important step into the problem has recently been achieved by Tian, Kamenev and 
Larkin~\cite{tian04}. They calculated the quantum corrections at time 
$t \gtrsim t_E$ where such corrections are small (but nonanalitic) in \kbar. 
From their approach it is clear that  the presence of $\phi$, by breaking 
the time reversal symmetry in the system, does affect quantitatively the quantum 
corrections to localization.

\subsection{Time reversal symmetry breaking -  COE to CUE crossover}
\label{coetocue}

Although the diagrammatic approach discussed in Ref.~\onlinecite{tian04} 
shades light on the role of the phase difference in the system, 
it cannot give quantitative predictions on the fully localized 
state at long time. We are interested in discussing the effect 
of $\phi$ in the localized state. The breaking of TRS by the superconducting phase 
difference, $\phi$, can be seen by direct verification that the Floquet operator in  
\eq{4evolution} at $\phi=0$ is invariant under the action of 
\be
        T: \left\{  \begin{array}{ccc}
        t & \raw& -t \\
        n & \raw& n \\
        \varphi & \raw& -\varphi \\
        \end{array}\right. \, ,
\ee
while such symmetry is broken at $\phi \neq 0$.

In the analogous problem of localization in disordered metals, it is known that 
the TRS breaking results in a reduction by a factor $1/2$ of the the weak localization 
corrections and in a doubling of the localization length (see Ref.~\onlinecite{rammer_book}).
The same kind of effects have been observed in chaotic systems as well~\cite{smilansky92}.
In fact, we also observe a doubling of localization length at $\phi=\pi/2$ 
with respect to the $\phi=0$ case as shown in \fig{localizationfig}. Note that the effect of 
doubling of the localization length has to be added to the effect of variation 
of the localization length through the change in the classical diffusion 
coefficient in  \eq{phase_effect}.

The effect of TRS breaking can be further investigated by looking at the distribution
of quasi-energies spacing. The reason behind this is the conjecture that 
quantum properties of classically chaotic systems are well described by the 
Random Matrix Theory (RMT)~\cite{bohigasconj}. This has been shown to 
hold for a wide class of chaotic systems~\cite{andreev96} (though exceptions exist~\cite{bohigas89}). 
In the RMT the breaking of the time reversal 
symmetry consists in a crossover from the Circular Orthogonal Ensemble (COE) 
to the Circular Unitary Ensemble (CUE) for the Floquet operator. 

\begin{figure}[t]
        \phantom{a}
        \begin{center}
        \includegraphics[width=0.45\textwidth]{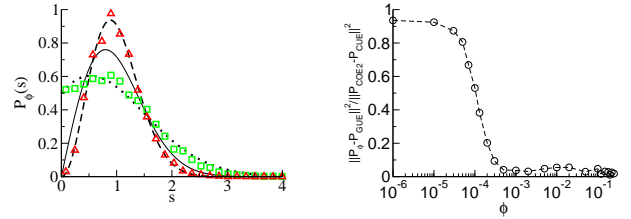}
        \end{center}
        \caption{(Color online). Left panel. Probability distribution for the quasi-energies 
        level spacing at $\phi=0$ (green squares) and 
        at $\phi=\pi/2$ (red triangles). The other curves are obtained 
        from RMT in case of COE (full line), ``$2$-folded'' COE (dotted line) 
        and CUE (dashed line). Right panel. 
        $||P_{\phi}(s)-P_{\trm{CUE}}(s)||^2/||P_{\trm{COE2}}(s)-
        P_{\trm{CUE}}(s)||^2$  as a function of 
        $\phi$. $||\cdot||$ is the $\mathcal{L}^2$ norm we use to characterize 
        the distance between curves. $P_{\trm{COE2}}(s)$ and 
        $P_{\trm{CUE}}(s)$ 
        are the probability distribution of level spacing 
        obtained from RMT for 
        ``$2$-folded'' COE and CUE respectively. $P_{\phi}$ is the probability
        distribution function obtained numerically at different 
        values of $\phi$.
        For both panel $k=200$ and $K=10$.}
\label{coe_cue}
\end{figure}

In order to check this conjecture we look at the level spacing probability distribution 
function $P(s)$ of the quasi-energies of the Floquet operator and compare them with the 
predictions of RMT. The results are presented in \fig{coe_cue} where our numerical results 
are shown together with the universal (no fit parameters) curves 
predicted by the random matrix theory. 

The plots have been obtained by considering a Floquet operator of $2^{10}$ 
levels averaged over $10$ realizations corresponding to random values of
$(\phi_L+\phi_R)/2$ distributed in $[0,2\pi)$.

For $\phi=0$ the probability distribution function is in perfect agreement with 
that of a folded spectrum~\cite{porter}. This corresponds to the fact that, 
at $\phi=0$, the Floquet operator of our system is $U(T,0)=F_1 \cdot  F_2 $, where 
$F_1$, $F_2$ are statistically independent Floquet operators of the usual QKR. 
The analysis of the transition between the two ensembles is presented in \fig{coe_cue}, 
right panel. We have analyzed the crossover between the two ensembles  by looking at the distance, at a  generic $\phi$,  
between the distribution $P_{\phi}$ ant the distribution 
corresponding to the Circular Unitary Ensemble, $P_{CUE}$.
The distance between the distributions is define  by the $L^2$ norm: 
$$
        || f(s)-g(s)||= \left[\int_{\mathbb{R}}(f(s)-g(s))^2 \, ds \right]^{1/2} \,\, .
$$
The results are plotted in \fig{coe_cue}, right panel. At $\phi=0$ the distribution 
coincides with that of a ``2-folded'' Circular Orthogonal Ensemble. At $\phi \sim 10^{-3}$ 
the distribution stabilizes to that of a a CUE. The crossover between the two ensembles 
is occurs at  $\phi \sim1/\sqrt{D}=\sqrt{2}/k \sim 5 \, 10^{-3}$.

\subsection{Transport properties in the chaotic regime}
\label{transport_chaotic}

Up to now we have discussed the chaotic dynamics by looking at the fluctuations of 
the charge on the central island. The possibility to observe the various fundamental aspects 
of quantum and classical chaotic behavior  in the Cooper pair shuttle has been 
discussed in Ref.~\onlinecite{romito05'}. We proceed further in this direction by investigating
how the chaotic dynamics affects the transport properties of the Cooper pair shuttle. 
The most convenient way to achieve our purpose is provided by the determination of the 
Full Counting Statistics (FCS) which was already analyzed in the normal shuttle by 
Pistolesi~\cite{pistolesi04} and in the charge regime of the Cooper pair shuttle 
by Romito and Nazarov~\cite{romito04}.
The generating function of the current cumulants 
at left and right contacts, defined through 
\begin{widetext}
\be
\left. \partial^n_{\lambda} \fk{Z}(\Lambda,\lambda,\tau)
\right|_{\lambda, \Lambda=0}=
\frac{(- \imath)^n}{e^n} \int_{\left[ 0, \tau \right]^{\times n}} \! \! \! \! 
dt_1 \cdot dt_n \,
{\lavg  \hat{I}(t_1) \cdots \hat{I}(t_n) \ravg}
\, \label{def_again}
\ee 
is expressed in terms of the Hamiltonian of the Cooper pair shuttle  $H_0$ in \eq{kicked_rotator} by
\bea
        & &  \fk{Z}(\lambda_L, \lambda_R \tau) = 
        e^{-\mathfrak{S}(\lambda_L, \lambda_R, \tau)} =
        \mathop{Tr} \underbrace{\left[\hat{\mathcal{U}}_{+\lambda_L, +
        \lambda_R}(\tau,0) \, \hat{\rho(0)} \,
        \hat{\mathcal{U}}^{\dag}_{-\lambda_L,-\lambda_R}(\tau,0) 
        \right]}_{\hat{\rho} ^{(\lambda_L, \lambda_R)} (\tau)} \, , 
\label{ale_fcs_general} \\
        &  & \hat{\mathcal{U}}_{\lambda_L,\lambda_R}(\tau,0)=
        \overrightarrow{T}exp^{-\imath \int_0^{\tau} \, 
        \hat{H}_{\lambda_L/2, \lambda_R/2}(s)\, \frac{ds}{\hbar} } \, , \,\, 
        \hat{H}_{\lambda_L/2, \lambda_R/2}=\hat{H}_0 (\phi_L \raw \phi_L -\lambda_L, 
        \phi_R \raw \phi_R -\lambda_R) \, .
\label{ale_fcs_general_1}
\eea
\end{widetext}
The necessary steps to calculate the FCS 
are summarized in Appendix~\ref{levitov}.
According to \eq{ale_fcs_general_1}, the task is to calculate the time evolution for 
$\rho^{(\lambda_L,\lambda_R)}$ 
which is a modified density matrix in which bra and ket evolves according to two 
different Hamiltonians. To this aim we  generalize the diagrammatic approach presented 
in Ref.~\onlinecite{altland93}. In the basis of charge eigenstates we write
\begin{widetext}
\bea
        \rho^{(\lambda_L,\lambda_R)}_{n_+,n_-}(t) &=& 
        \sum_{n_+',n_-'} G^{(\lambda_L,\lambda_R)}_{n_+,n_+';n_-,n_-'}(t) 
        \rho_{n_+',n_-'}(0) \, , \label{4_fcs_density} \\
        G^{\lambda_L,\lambda_R}_{n_+,n_+';n_-,n_-'}(t) &=&
        \Avg{n_+ | \hat{V}\hat{U}_{+\lambda_L, +\lambda_R}^t | n_+'} 
        \Avg{n_-' | \hat{U}_{-\lambda_L, -\lambda_R}^{\dag t} 
        \hat{V}^{\dag}| n_-} \, ; \label{4_kernel}
\eea
$t$ is the time counted in integers multiples 
of the period $T$ and $U_{\lambda_L, \lambda_R}$ is given in \eq{4evolution} after the 
replacing $\phi_L \raw \phi_L - \lambda_L$, $\phi_R \raw \phi_R - \lambda_R$. 
In \eq{4_kernel} we have also added an  operator for the free evolution after 
the last kick for computational convenience; it does not affect the physics because it 
simply means that we look at all observables just before a kick, rather than just after it.
We now turn to the Wigner representation for the kernel $G$, 
\be
\tilde{G}^{(\lambda_L,\lambda_R)}_{n,\Theta;n',\Theta'}(t)=
\sum_{n_+-n_-}\sum_{n_+'-n_-'} G^{(\lambda_L,\lambda_R)}_{n_+,n_+';n_-,n_-'}(t) 
e^{i\Theta(n_+ - n_-)} e^{i\Theta'(n_+' - n_-')} \, ,
\label{4_wigner}
\ee
\end{widetext}
where $n=(n_++n_-)/2$ and $n'=(n_+'+n_-')/2$ are  the ``center of mass'' coordinates.
We compare the kernel in Wigner representation with 
the classical equivalent kernel for the evolution 
of a distribution function in the phase space.
The classical dynamics is unstable in the $\Theta$ 
direction~\cite{izrailev90}, and therefore we approximate the quantum 
kernel $\tilde{G}$ by its average over $\Theta$, $\Theta'$.
As a consequence the inverse Fourier transform of \eq{4_wigner}
picks up non-vanishing terms only at $n_+=n_-$ and
$n_+'=n_-'$. 
In doing so, we have reduced \eq{4_fcs_density} to an evolution equation for the populations 
of the modified density matrix,
\be
        \rho^{(\lambda_L,\lambda_R)}_{n,n}(t)  = 
        \sum_{n'}  G^{(\lambda_L,\lambda_R)} (n,n';t) \rho^{(\lambda_L,\lambda_R)}_{n',n'}(0) \, , 
\label{4_good_kernel} 
\ee
\bea
        G^{(\lambda_L,\lambda_R)}(n,n';t) & = & 
        \Avg{n | \hat{V}\hat{U}_{+\lambda_L,+\lambda_R}^t | n'} \cdot \nonumber \\
        && \Avg{n' | \hat{U}_{-\lambda_L,-\lambda_R}^{\dag t} 
        \hat{V}^{\dag}| n} \, , 
\label{4_good_kernel_2}
\eea
and therefore
\bea
        \fk{Z}(\lambda_L,\lambda_R,t) &=& \mbox{Tr}\{ \rho^{(\lambda_L,\lambda_R)} (t) \}
        \nonumber \\
        &=&
        \sum_{n,n'}G^{(\lambda_L,\lambda_R)}(n,n';t) \rho_{n',n'} (0) \, ,
\label{4_intermediate}
\eea
We expect that $G^{(\lambda_L,\lambda_R)}(n,n';t)$ depends only on the difference $n-n'$, 
and we then average out $(n+n')/2$ as follows
\be
        \overline{G}^{(\lambda_L,\lambda_R)}(n-n';t) \! 
        \equiv \!  \lim_{L_0 \raw \infty} 
        \sum_{l=-L_0}^{L_0} 
\frac{G^{(\lambda_L,\lambda_R)}(n+l,n'+l;t)}{2L_0+1} \, .
\label{4_averaging}
\ee
The key observation to construct the perturbation theory
is that the average of the kernel can be reabsorbed in  
an average over the free evolution operators by means
of the relation
\be
\Avg{n+l | \hat{V} | n'+l} =
\Avg{n | \hat{W} | n'} \, ,
\label{4_randomv}
\ee
with
\be
\Avg{n | \hat{W} | n'}=
\exp\left[-i \kbar (n-l)^2/2\right] \delta_{n,n'} \equiv W(n) \delta_{n,n'} .
\label{4_trick}
\ee
The kernel $G^{(\lambda_L,\lambda_R)}(n-n';t)$ 
involves products with an equal number of $\hat{W}$ 
and $\hat{W}^{\dag}$, therefore the average 
is completely defined by 
\bea
&  & \overline{W(n_1) W(n_2)  \dots W(n_i) 
W(n_1')^{\ast} W(n_2')^{\ast} \dots W(n_i')^{\ast}} = \nonumber \\
& & = \exp \left[-(i \, \kbar/2)
\sum_a (n_a^2-{n_a'}^{2}) \right] 
\delta_{\sum_a (n_a-n_a',0)}\, .
\label{4_basta}
\eea
The potential $\hat{W}$ of the kicked rotator is not 
Gaussian distributed. 
However, in the diagrammatic expansion we are going to perform, 
we will restrict 
only to two point correlation functions. 
Therefore the r.h.s. of \eq{4_basta} reduces to 
$\delta_{n,n'}$ and the effects of having a 
non-Gaussian distribution  don't appear~\cite{altland93}.
Ones the kernel $G^{(\lambda_L,\lambda_R)}(n,n';t)$ 
in \eq{4_intermediate} has been replaced by the averaged one
which depends only on the difference $n-n'$, 
the sum over $n'$ in the same equation can be 
performed noting that $\sum_{n'}\rho_{n',n'} (0)=
\mathop{Tr} \rho(0) =1$.
We now turn to phase and frequency space
by 
Fourier transforming the kernel $\overline{G}^{(\lambda_L,\lambda_R)}(n-n';t)$ 
both in $n$, $n'$ and in $t$, obtaining
\be
\fk{Z}(\lambda_L,\lambda_R,t)=\lim_{q\raw 0} 
\int_0^{2 \pi} e^{-i \omega T t} 
\overline{\fk{G}}^{(\lambda_L,\lambda_R)}(q;\omega) \, .
\label{4_nice_fcs}
\ee
The kernel $\fk{G}$ is defined through
\be
\overline{\mathcal{G}}^{(\lambda_L,\lambda_R)}(\theta,\theta',q,q';\omega,\omega_0)=
\overline{\fk{G}}^{(\lambda_L,\lambda_R)}(q;\omega) \delta(q-q') \, .
\label{4_boring_def}
\ee
with
\begin{widetext}
\be
        \overline{\mathcal{G}}^{(\lambda_L,\lambda_R)}(\theta,\theta',q,q';\omega,\omega_0)
        = \overline{\Avg{\theta_+ | \hat{W}\left( 1- e^{i\omega_+ T}
        \hat{\mathcal{U}}_{(+\lambda_L,+\lambda_R)}\right)^{-1} | \theta'_+}
        \Avg{\theta_-' |  \left( 1-
        e^{-i\omega_- T} \hat{\mathcal{U}}^{\dag}_{(-\lambda_L,-\lambda_R)} 
        \right)^{-1} \hat{W}^{\dag} | \theta_-}} 
\, , 
\label{4_perturb_start}
\ee
\end{widetext}
 $\theta_{\pm}=\theta\pm q/2$, 
$\theta'_{\pm}=\theta'\pm q'/2$, $\omega_{\pm}=\omega_0\pm \omega/2$ and  
$\hat{\mathcal{U}}_{\lambda_L, \lambda_R}=\hat{J}_R\hat{W}\hat{J}_L\hat{W}$. 
The dependence of $\hat{\mathcal{U}}_{(\lambda_L, \lambda_R)}$ 
on $\lambda_{L(R)}$ is through
$\hat{J}_{L,R} = e^{-i k \cos(\hat{\varphi}- (\phi_{L,R}+\lambda_{L,R}))}$,
$\hat{J}^{\dag}_{L,R} = e^{i k cos(\hat{\varphi}- (\phi_{L,R} -\lambda_{L,R}))}$.
The diagrammatic perturbation theory consists in a series   
expansion of the r.h.s. of \eq{4_perturb_start} 
in $\hat{J}_R\hat{W}\hat{J}_L\hat{W}$ . 
Any term of the expansion 
correspond to a diagram which consists of propagators,
\be
\begin{array}{c}
        \includegraphics[height=1.5cm]{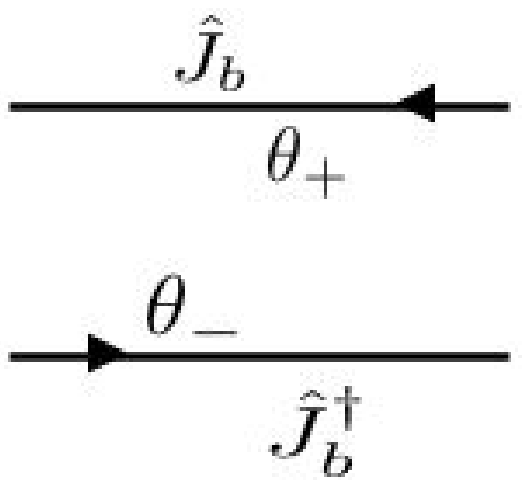}\end{array} 
        \!\!
        \begin{array}{c}= e^{i\omega T}e^{-ik\cos (\theta+\frac{q}{2}-\phi_b-\lambda_b)}
        e^{ik\cos (\theta-\frac{q}{2}-\phi_b+\lambda_b)}\end{array}  .
\label{4_propag}
\ee
and averaged vertices 
\be
\begin{array}{c}
\includegraphics[height=1.5cm]{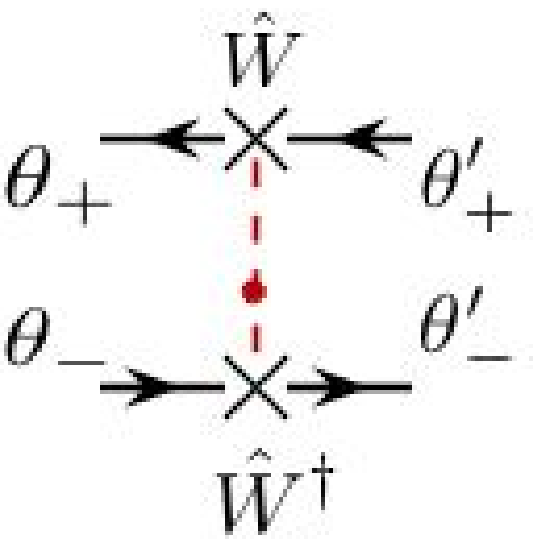}\end{array} 
\begin{array}{c}= \delta(\theta_+'-\theta_+ -(\theta_-'-\theta_-)) \, .
\end{array} 
\label{4_vertices}
\ee
We have absorbed the frequency dependence in the 
expression for the propagator, from which it is evident that 
the final expression for the kernel is independent on $\omega_0$.
The average process is indicated by a dotted line in the diagrammatic
expression. In any diagrams, correlations can 
take place between any pair of vertices one of which in the 
upper line and the other in the lower line. Any other term involves
expressions of the form $\overline{\hat{W}\hat{W}}$ or 
$\overline{\hat{W}^{\dag}\hat{W}^{\dag}}$ and vanishes accordingly
to \eq{4_basta}.
 
 For $k \gg 1$, 
diagrams involving crossing correlations lines, 
are parametric
small in $1/k$, {\it i.e.} in $\kbar$. 
This allows us to approximate, at lowest order in $\kbar$, 
which is the classical limit, 
the kernel with the sum of all ladder diagrams, 
we will refer to as {\em diffuson}.

\subsection*{Classical limit}

The diffuson is defined through the series of ladder diagrams
\begin{widetext}
\be
\begin{array}{c}\includegraphics[height=1.5cm]{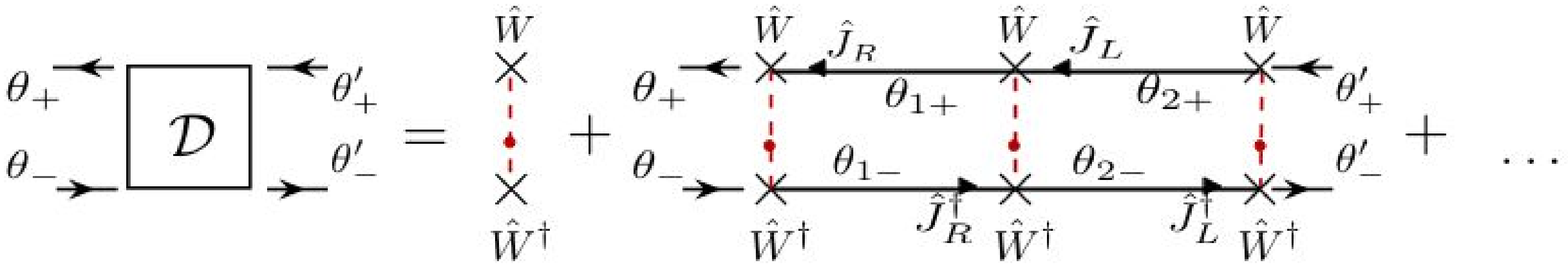}
\end{array} \, .
\label{4_diffuson_1}
\ee
The computation of the diffuson become simple 
due to \eq{4_vertices}.
It allows to factorize any diagram as products of 
\be
\begin{array}{c}
\zeta(q;\omega)
\end{array}=\begin{array}{c}
\includegraphics[height=1.5cm]{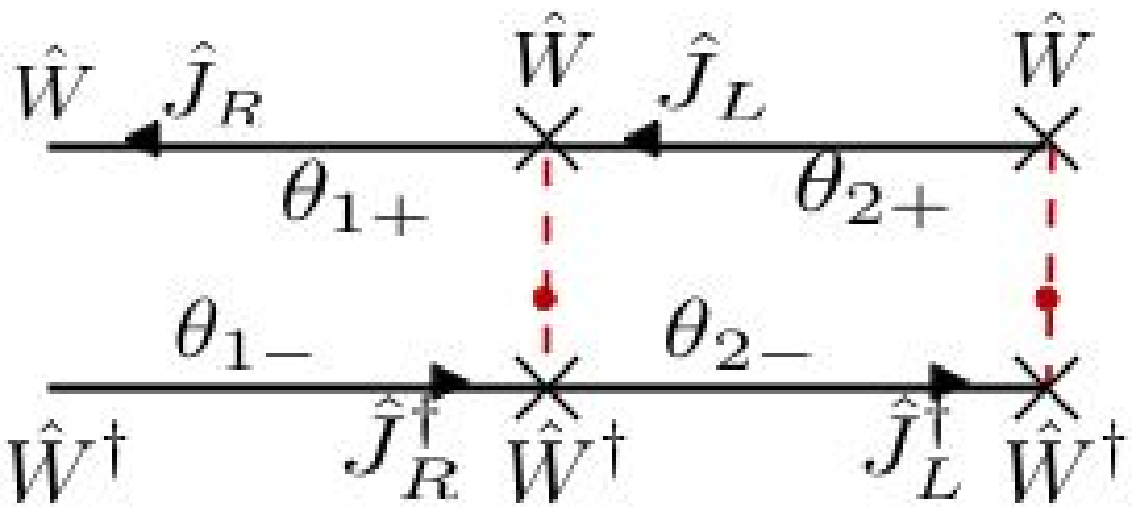}
\end{array}=\begin{array}{c}
 J_0(2k\sin(q/2-\lambda_L)) J_0(2k\sin(q/2-\lambda_R)) 
 e^{i\omega T}
\end{array} \, .
\label{4_zeta}
\ee 
then the series is a geometric series and we find
\be
\mathcal{D}^{(\lambda_L,\lambda_R)}
(q,q';\omega)= \delta(q-q') 
\sum_{i=0}^{\infty}[\zeta(q;\omega)]^i=\frac{1}{1-e^{i\omega T}
 J_0(2k\sin(q/2-\lambda_L)) J_0(2k\sin(q/2-\lambda_R))} 
 \, \delta(q-q') \, .
\label{4_fin_diff}
\ee
By replacing $\overline{\fk{G}}^{(\lambda_L, \lambda_R)}
(q,\omega)$ with $\mathcal{D}^{(\lambda_L,\lambda_R)}
(q,q';\omega)$ in \eq{4_nice_fcs}, 
 we finally find
\be
\fk{Z}(\lambda_L, \lambda_R,t)= e^{-\fk{S}(\lambda_L, \lambda_R)t}= \left[ J_0(2k\sin \lambda_L)
 J_0(2k\sin \lambda_R) \right]^t \, .
\label{4_fcs_classical}
\ee
\end{widetext}
It defines all transport properties of the system.
In particular the probability of charge transfer per cycle 
at a given contact is
\bea
        p_N^{L(R)} &=&\lim_{\lambda_R, \lambda_L \raw0} 
        \frac{1}{2 \pi} \int_{0}^{2 \pi} d\lambda_{L(R)} 
        e^{\fk{S}(\lambda_L, \lambda_R)} 
        e^{- i \lambda_{L(R)} N} 
        \nonumber \\
        &=& \sum_{M \in \mathbb{Z}} 
        \delta_{N, 2M} J_{M}^2(k) \, .
\label{4_probab}
\eea
The properties of Bessel functions guarantees the normalization 
condition $\sum_N p_N =1$. Probabilities 
are non-vanishing only for even numbers of electrons, 
i.e. Cooper pairs. Note also that the very same existence of such
probabilities means that the transport is completely incoherent 
and the chaotic dynamics is efficient in destroying 
any features of the coherent superconducting nature of 
the transfer. 

 The limit $\lambda_{L,R} \raw 0$ of \eq{4_fin_diff} 
 gives us the kernel for the evolution of the density matrix 
 in the classical limit. 
 In particular, for long time and large $n-n'$ 
 ($\omega \raw 0$, $q\raw 0$),
\be
\mathcal{D}(q,q';\omega) \approx \frac{1}{(kq)^2/2-i\omega T t} 
\delta(q-q') \, .
\label{4_long_large_limit}
\ee
which expresses the diffusive behavior 
in $n-n'$ with diffusion constant $D=k^2/2$
in complete agreement with the results obtained from the classical map.  It correctly reproduces the 
result obtained in  Ref.~\onlinecite{altland93}. 
The dependence on $\phi$ disappears 
from the final expression. 
In fact we do not recover the classical effects 
in \eq{phase_effect} 
because we just restrict our considerations to single impurity 
correlations, while the dependence 
on $\phi$ comes from higher time correlations 
(see Appendix~\ref{diffusion_corrections}). 
It is possible to reproduce the classical 
corrections to the diffusion constant within the 
diagrammatic approach. This has been performed in 
Ref.~\onlinecite{tian05} for $\phi=0$.

\subsection*{Beyond the classical limit}

The calculation of quantum corrections to the classical limit of the FCS 
in the diagrammatic approach is based on writing a Dyson equation 
for the four point vertex $\overline{\mathcal{G}}^{(\lambda_L,\lambda_R)}(\theta,\theta',q,q';\omega,\omega_0)$, as discussed by Altland~\cite{altland93} in the case of $\lambda_L=\lambda_R=0$. 
In the limiting case 
$\lambda_L=\lambda_R=0$ \eq{4_fcs_density} describes 
the time evolution of the density matrix of the system and, 
due to probability conservation, 
the leading quantum corrections at $\phi=0$ are determined only 
by the most infrared divergent term of the diagrammatic expansion, the cooperon~\cite{altland93}. 

The cooperon consists of the the sum of maximally crossed diagrams and, in the general case $\lambda_L,\lambda_R \neq 0$, is defined by the diagrams series
\begin{widetext}
\be
\begin{array}{c}\includegraphics[height=1.5cm]{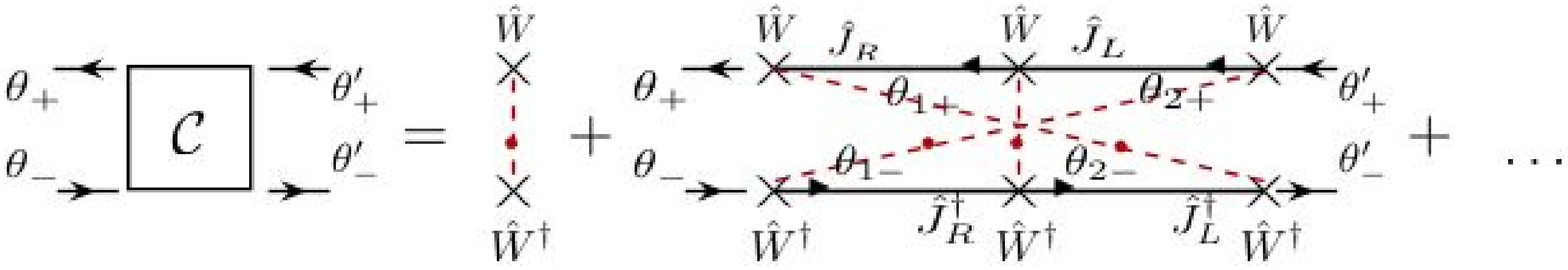}
\end{array} \, .
\label{4_cooperon_1}
\ee
The cooperon diagram is topologically 
equivalent to a diffuson diagram after time reversion of the lower line,
\be
\begin{array}{c}\includegraphics[height=1.5cm]{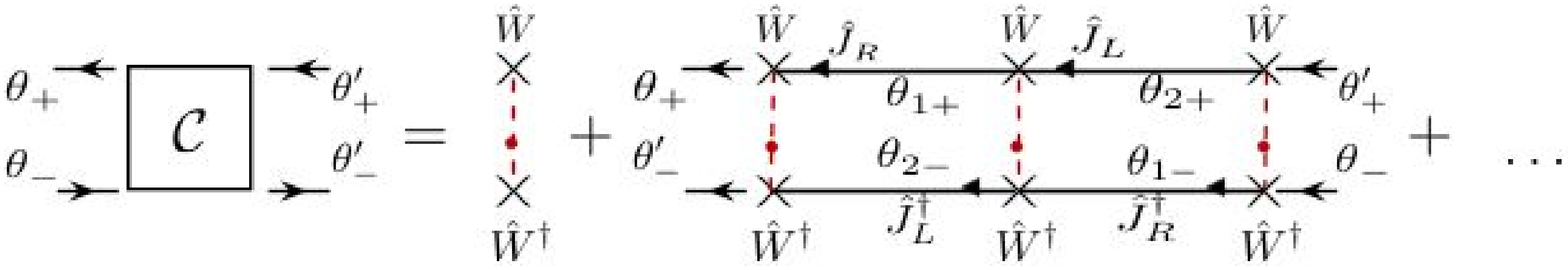}
\end{array} \, .
\label{4_cooperon_2}
\ee
In this form the correlation lines do not cross and it is 
clear that the contribution of the cooperon series is at same order of the diffuson's one in the $1/\kbar$ expansion.
The sum of the diagrams series in \eq{4_cooperon_2} 
leads to
\be
\mathcal{C}(\vartheta,\vartheta';\omega)=\frac{1}{1-e^{i\omega T}
 J_0(2k\sin(\vartheta/2-(\lambda_L/2-\lambda_R/2))) J_0(2k\sin(\vartheta/2+\phi-(\lambda_R/2-\lambda_L/2)))} \, 
 \delta(\vartheta-\vartheta') \, ,
\label{4_fin_coop}
\ee
\end{widetext}
with $\vartheta=\theta_++\theta_-'$, $\vartheta'=\theta_+'+\theta_-$.

If $\lambda_L=\lambda_R=0$ the cooperon in the low frequency, small angle limit becomes
\be
\mathcal{C}(\vartheta,\vartheta';\omega) \approx 
\frac{1}{k^2/4 \left[ \vartheta^2 +(\vartheta+2\phi)^2\right] -i\omega T t} 
\delta(\vartheta-\vartheta') \, .
\label{4_long_large_coop}
\ee
At $\phi=0$ the cooperon is divergent in the infrared limit,  
$\omega \ra 0, \theta \ra 0$,
and the quantum corrections can be computed.
This is not the case at $\phi \neq 0$ (form 
\eq{4_long_large_coop}, $\lim_{\omega \ra 0, \theta \ra 0}
\mathcal{C}(\vartheta,\vartheta';\omega)$ is finite) 
where, due to the breaking of time reversal symmetry, the
diagrammatic approach does not work~\cite{altland93}.
In fact, by Fourier transforming the previous equation 
with respect to time and phase 
it can be easily seen that the diffusive behavior 
is modified by the presence of $\phi$ introducing 
a relaxation time $\tau_{\phi}=[(k\phi)^2/2]$.
At $\tau_{\phi}\sim 1$ , which is 
$\phi \sim 1/\sqrt{D}$, the cooperon contribution is 
completely suppressed. 
 
The described procedure cannot be generalized to include
non-vanishing counting fields to determine 
the quantum corrections to the classical result for the FCS. 
In fact such a procedure for $\lambda_L=\lambda_R=0$ crucially 
relays the possibility of writing a Dyson equation for the 
vertex function and ultimately on the existence 
of a Ward identity corresponding to probability 
conservation~\cite{altland93}, $\trm{Tr}\{ \rho(t)\}=1$.
In the generalized case, due to presence of the counting fields 
in the kernel 
$G^{(\lambda_L,\lambda_R)}(n,n';t)$,
$\mathop{Tr}\{ \rho^{(\lambda_L,\lambda_R)} \} \neq 1$, 
a similar conservation law cannot be established when the 
counting fields are inserted in the propagators. 
Let us note that, even if we had an explicit expression for the quantum corrections to the classical FCS, it would coincide with the result of Ref.~\onlinecite{altland93} in the limit $\lambda_L=\lambda_R=\phi=0$.
In this specific limit the quantum corrections are shown to be divergent thus implying that they cannot be treated 
as a perturbation of the classical limit. 
Therefore such a perturbative calculation
cannot give us quantitative information on the quantum steady state 
of the system~\cite{altland93}, i.e. the dynamical localized state.

Anyway our results for the FCS give us direct predictions for the 
current and current noise in the system. 
From \eq{4_fcs_classical} we find $\Avg{I_L}=\Avg{I_R}=0$.
The low frequency noise at various contact is easily determined
\be
S_L(\omega=0)=S_R(\omega=0)=k^2/4 \, .
\label{4_noise}
\ee
showing a signature of the diffusion of the wavefunction  
in the charge space.
The correlation between left and right currents is vanishing:
\be
\int_{\mathbb{R}}\avg{I_L(0)I_R(t)} =0 \, .
\label{correlations}
\ee
These results are valid in the classical limit $\kbar \rightarrow 0$. 
Realistic values of the parameter can range from $(K \sim 10,k \sim 15)$ to
$(K \gtrsim 30, k \sim 10^3))$.
In any case with a finite non-vanishing $\kbar$, after a time $t^{\star}\sim T K^2$,
the system reaches a steady state with a localized wavefunction.
In this regime we can easily estimate the average value of the 
current and the fluctuations of the current signal. This is 
achieved assuming random uncorrelated phases for the 
localized wave function of the grain:
\bea
& & \ket{\psi} = \frac{1}{\sqrt{2 l+1}}\sum_{n=-l}^{l} 
e^{i \alpha_n} \ket{n} \, ,\label{localized} \\
& & P(\alpha_n)=1/(2 \pi) \, \,\,\,\, \alpha_n \in [0,2\pi) \, . \label{non_serve}
\eea
It follows that 
\bea
\overline{I_{L(R)}} &= & 0 \, , \label{cur_1}\\
\sqrt{\overline{I_{L(R)}^2}} &=& \frac{2e}{T\sqrt{2l+1}} \, ,
\label{cur_2}
\eea
where the $\overline{\cdot}$
defines (only in \eqs{cur_1}{cur_2}) 
the average over the random phases of the 
usual Josephson current 
$\hat{I}_{L(R)} =2e \frac{k}{T}\avg{\sin (\hat{\varphi} - \phi_{L(R)})}$
defined in \eqs{current_L}{current_R}.
Also in the localized state the current is suppressed 
by the chaotic dynamics while the typical value of 
the current fluctuations is an indirect measurement of the 
localization length.

\section{Conclusions}
\label{conclusions}

The Cooper pair shuttle is a very rich system where to study various 
aspects of coherence quantum dynamics of driven systems. In this work 
we concentrated on two different regimes where the charging energy is 
either much smaller or much larger than the Josephson coupling energy.
In both cases we analyzed the transport properties as the Josephson 
current or the current fluctuations. In the charge regime we evaluated 
both the DC and AC Josephson effect. In the other regime we analyzed 
the consequences of the underlying classical chaotic dynamics on the full 
counting statistics.

We believe that the possibility to realize periodic driving via time-dependent 
fluxes opens the possibility to study the very exciting area of 
quantum driven systems in variety of different situations beyond that
considered in this paper.

\acknowledgments
We acknowledge useful discussions  with F. Plastina, G. Benenti, A. Kamenev, and 
Yu. V. Nazarov. This work was supported by EC through grant EUROSQIP and by MIUR 
through PRIN.

\appendix

\section{Effect of non-degeneracy during the cycle}

\label{voltaggio_costante}

In this Appendix we discuss the Josephson 
current in the Cooper pair shuttle when the gate voltage fluctuates around a 
fixed value away from the degeneracy point even during the connection to 
the electrodes. In Section \ref{calcoli} we assumed $n_g(t)=1/2$ during the Josephson 
contacts to enhance the Cooper pair transfer and $n_g(t) =\trm{const.} \neq 0$ 
during the free evolution time. From an experimental point of view, however, 
it would be easier to avoid this periodical modification of 
the gate voltage $V_g$. It can be therefore interesting to have an expression for the 
DC Josephson current in the case of constant gate voltage.

If $n_g(t)=\trm{const.}=1/2$, the Josephson current can be 
obtained from the expressions of section~\ref{calcoli} 
with the replacement $\chi_{\ra}=\chi_{\la}=0$.
If $n_g(t)=\trm{const.} \neq 1/2$, instead, the general 
expression for for the current (Eq.~(\ref{formal_current})) 
still holds, but differences arise in the explicit form 
of matrix $M_0$ and vector $\mathbf{v}_0$.
During the free evolution time intervals the dynamics is 
unchanged compared to the case discussed before. 
In the L and R regions instead, we have to include in the Hamiltonian 
the term proportional to the charging energy difference, $E_C$,  between the
two charge states. T Hamiltonian (system and bath), 
in the basis which diagonalizes the Cooper pair box Hamiltonian now reads 
\be
        H= \frac{E}{2} \sigma_z +\hat{\mathcal{O}}(\cos(2 \mu)\sigma_z 
        +\sin(2\mu) \sigma_x)+H_{bath} \;,
\label{h_E_C}
\ee
where $\mathcal{O}$ is the same of \eq{hi}, and
\be
E=(E_C^2+E_J^2)^{1/2} \,\,\, , \,\,\,\,\, \cos(2\mu)=E_C/E \nonumber \, .
\ee
The Hamiltonian in Eq.~(\ref{h_E_C}) has been widely studied~\cite{weiss99}.
In Born-Markov and Rotating Wave approximations the time evolution of 
populations (diagonal terms) in the reduced density matrix and coherences 
(off-diagonal ones) are still independent. 
The respective decoherence rates read
\bea
\gamma_J^{\trm{pop.}} & =& 2\gamma_J \sin^2(2\mu) \; ,\\
\gamma_J^{\trm{coher.}} & = & \gamma_J \sin^2(2\mu) + \Gamma_J \cos(2\mu) \; .
\eea
We notice that we have to introduce two different 
dephasing rates, $\gamma_J$ and $\Gamma_J$. In our approximation, they are
$\gamma_J=(\pi/2) \alpha E \coth(E/T_b)$, and 
$\Gamma_J=2 \pi \alpha T_b$, in case of weak coupling 
of the system to the bath ($\alpha \ll 1$).
Depending on the relative strength of the two energy scales
$E_C$ and $E_J$, we have different effects.
If $E_C=0$, (corresponding to $2\mu=\pi/2$) we recover 
the Hamiltonian of the early case described in 
section~\ref{calcoli}.
If $E_C \ll E_J$, we have corrections of order $E_C/E_J$ 
in our previous results, and we are not interested in them
as we get a finite result at zero order in $E_C/E_J$.
In the opposite limit $E_J \ll E_C$, the situation is quite different.
If one neglects $E_J$, the current vanishes because $\hat{n}$ 
is a constant of motion. The first non vanishing term must 
be of order $E_J/E_C$.
Below we give the analytical expression for the current in the limit
of strong dephasing (for $t_L=t_R=t_J$ and $t_{\ra}=t_{\la}=t_C$) including only the 
leading order in $E_J/E_C$. The strong dephasing limit refers to 
the condition $\gamma_J t_J, \Gamma_J t_J, \gamma_C t_C \gg 1$, 
which allows a series expansion of the Josephson current at first order 
in $e^{-\gamma_J t_J}$, $e^{-\Gamma_J t_J}$ and $e^{-\gamma_C t_C}$:
\begin{widetext}
\be
        I_{strong}  \sim  - \frac{2 e}{T}
        \tanh \left( \frac{E_C}{T_b} \right) \left(\frac{E_J}{E_C} \right)^2 
        \sin \phi \, e^{-\gamma_C t_C}
        \left[ \sin(2\chi) 
        \left( \cos (2 \theta) e^{-\gamma^{\trm{coher.}}t_J}
        -e^{-\gamma^{\trm{pop.}}t_J} \right) + \cos(2 \chi) 
        e^{-\gamma^{\trm{coher.}}t_J} \right] \, .
\label{current_E_C}
\ee
\end{widetext}
In this cases the Josephson energy does not any more enter the current 
through the combination $E_J t_J$ as in the previous case, but rather it 
appears through the ratio $E_J/E_C$. There is an overall suppression 
factor $\propto (E_J/E_C)^2$. 
Note that the dependence on the dynamical phase $\chi$
is not the same of that in Eq.~(\ref{limit0}).
The approximation $E_J \ll E_C$ also induces a hierarchy in the 
dephasing rates, $
\gamma_J^{\trm{coher.}} \gg  \gamma_J^{\trm{pop.}} \sim 2 
\left( E_J/E_C \right)^2 \gamma_J $.
Within such approximation, we can neglect terms of order 
$e^{-\Gamma_J t_J}$ with respect to $1$ (or equivalently 
$e^{-\gamma^{\trm{coher.}}t_J}$ with respect to $e^{-\gamma^{\trm{pop.}}t_J}$)
in the current expression, leading to
\begin{widetext}
\be
I_{strong} \sim  - \frac{2 e}{T}
\tanh \left( \frac{E_C}{T_b} \right) \left(\frac{E_J}{E_C} \right)^2 
e^{-\left( 2 \left( \frac{E_J}{E_C} \right)^2 \gamma_J t_J+\gamma_C t_C\right)}
\sin(2\chi) \, \sin\phi \, .
\label{current_E_C'}
\ee
\end{widetext}
The exponential suppression due to dephasing 
in the Josephson contact times is reduced by the factor 
$\sin(2\mu) \sim (E_J/E_C)^2$.

\section{Phase dependent corrections to the classical diffusion}
\label{diffusion_corrections}

In this Appendix we derive~\eq{phase_effect} for the charge diffusion coefficient 
obtained for the modified Chirikov map 
in~\eq{classical_map}. 
The starting point of the calculation, 
following the idea of Ref.~\onlinecite{rechester81}, 
is to rewrite the~\eq{classical_map},
\be
\left\{ \begin{array}{ccl}
p_t & = & p_{t-1}- K \sin\left\{\theta_{t-1} - [(t+1)\mbox{~\small mod~} 2]\phi \right\}\\
\theta_t & = & \theta_{t-1}+ p_t, 
\end{array} \right. \, ,
\label{c_1}
\ee
 as a time 
evolution equation for the probability distribution function 
in the phase space, $P(\theta,n;t)$ in which 
we add a further random step. It reads
\begin{widetext}
\be
P(\theta,n;t)=\int_0^{2 \pi} G(\theta-\theta', n) \,
P \left( \theta', n+K \sin \left\{ \theta' - 
[(t+1)\mbox{~\small mod~} 2]\phi \right\}; t-1 \right) \, dx \, ,
\label{c_2}
\ee
where $t$ counts the number of kicks.
The kernel $G$ defines the random step  
between two kicks,
\be
G(\delta \theta, n)=\frac{1}{\sqrt{2 \pi \sigma}} 
\sum_{m=-\infty}^{+\infty} 
\exp \left[ - \frac{\left( \delta \theta - n +2 \pi m\right)^2}{2 \sigma} \right]
=\frac{1}{2 \pi} \sum_{m=-\infty}^{\infty} e^{-\sigma m^2/2} 
e^{i m (\delta \theta -n)} \, .
\label{c_3}
\ee
\end{widetext}
It correspond to add a diffusive term 
($\sigma/2$ the diffusion coefficient) 
in the differential equation for the time evolution  
between two kicks. 
In this way we have replaced the deterministic dynamics 
of \eq{c_1} with a stochastic one which reproduce the effect of chaotic dynamics.
At the end of the calculations we can take the 
limit $\sigma \raw 0$.

By introducing the Fourier coefficients for the 
probability distribution function $P(\theta, n;t)$ defined through
\be
P(\theta,n;t)=\frac{1}{(2 \pi)^2} \sum_{m=-\infty}^{\infty} 
\int_{\mathbb{R}} dp \, a_m^{(t)}(p) \, e^{i (m\theta +pn)} \, ,
\label{c_4} 
\ee
we can write the diffusion coefficient as
\be
D=\frac{1}{2t} \avg{n^2} =\frac{1}{2t} \lim_{p \raw 0^+}
\left( i \frac{\partial}{\partial p} \right)^2 a_0^{(t)} (p) \, .
\label{c_5}
\ee
We have implicitly assumed that $p_0 =0$.
We now rewrite the time evolution in~\eq{c_2} in the Fourier 
space and then we expand in powers of $1/\sqrt{K}$.
The first step is performed by simply inserting \eqs{c_3}{c_4} 
into \eq{c_2}. After some algebra, we find
\be
        a_m^{(2t)}(p) = \left[ \sum_{l=-\infty}^{+\infty} 
        J_l(K|p'|) \, e^{i \phi \mathop{sgn}(p')}\,
        e^{-\sigma m^2/2} \right] a_{m'}^{(2t-1)}(p')  
\label{c_6} 
\ee
\bea
        a_m^{(2t+1)}(p) &=& \! \!\left[ \sum_{l=-\infty}^{+\infty} 
        J_l(K|p'|) \,
        e^{-\sigma m^2/2} \right] a_{m'}^{(2t)}(p')  \label{c_7} \\
        & & p'=p+m  \,\, \, \,\,\, m'=m-l \mathop{sgn}(p') \, ,
\label{c_8} 
\eea
where $J_l(x)$ is the $l$th Bessel function. 

If we represent the variables $p$ and $m$ respectively in the 
$x$-axis and $y$-axis in the plane, \eq{c_8} defines a path 
in such a plane. The calculations of the diffusion constant 
through \eq{c_5} is therefore reduced to the calculation
of $a_m^{(t)}(p)$ along the path defined by \eq{c_8}.
Indeed many path can contribute and one has to sum over them. 
As it is evident from \eq{c_5} only terms of $a_m^{(t)}(p)$ 
linear in time are important, therefore we can consider an even number of periods.
We assume the initial condition 
$a_m^{(t=0)}(p)=\delta_{m,0}\delta_{p,0^+}$. Then, from \eq{c_8}
the first step of the path is either $(0,0) \raw (-1,1)$, or 
$(0,0) \raw (1,-1)$, or $(0,0) \raw (0,0)$.
From \eq{c_5} we realize that also the final point of the path
have to be at $m=0$. 
The number of steps of the path correspond to 
the number of kicks.
A trivial path consists in remaining at the origin,
\be
\underbrace{(0^+,0) \raw (0^+,0) \raw \cdots \raw (0^+,0)}_{2t+1} \, .
\label{c_9}
\ee
The contribution of this path is  
\begin{widetext}
\be
a_0^{(2t)}(p\raw 0^+) = \left[ J_0(K|p|) \right]^{2t} a_0^{(0)}(p)
\approx \left[ 1-2t \left(\frac{Kp}{2}\right)^2 \right] a_0^{(0)}(0) \, ,
\label{c_10}
\ee
\end{widetext}
which, combined with \eq{c_5}, gives $D=K^2/4$.
This is the standard result corresponding to the 
zero order term in $1/\sqrt{K}$ expansion.
Any other term corresponds to a path with
 steps moving away from the origin. 
 Any such step in the maps 
in \eqs{c_6}{c_7} involves
a Bessel function $J(x\sim K)$, which, in the limit $K \gg 1$  decays 
like $\sim 1/\sqrt{K}$. Therefore a perturbation in $1/K$ is a 
perturbation in the number of steps away from the origin 
in the path. 
The path in \eq{c_9} is the only contribution at zero order and it is independent 
on the phase difference $\phi$.

The first correction to the result in \eq{c_10} involves 
the path $(0,0) \raw (1,-1) \raw (0,1) \raw (0,0)$ and 
its symmetric with respect to the origin of axes. 
By identifying the path with its 
numerical value, we write the 
corrections to the diffusion constant as
$\delta D = \mathcal{P}_1 +\mathcal{P}_2$, with 
\be
\mathcal{P}_1  =  \begin{array}{c}
\includegraphics[width=0.16\textwidth]{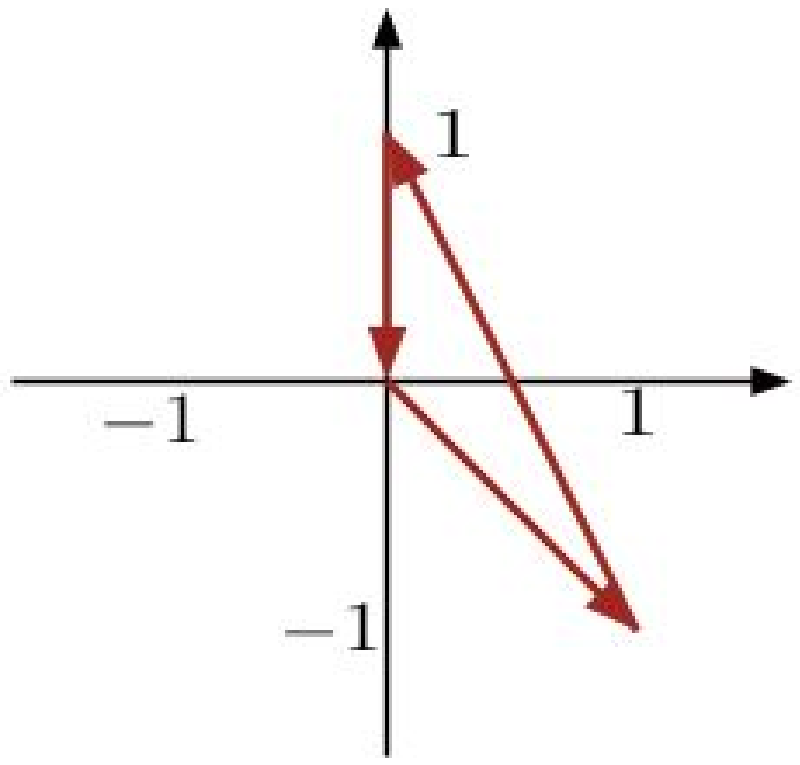} 
\end{array} \, , \,\,
\mathcal{P}_2  = 
\begin{array}{c}
\includegraphics[width=0.16\textwidth]{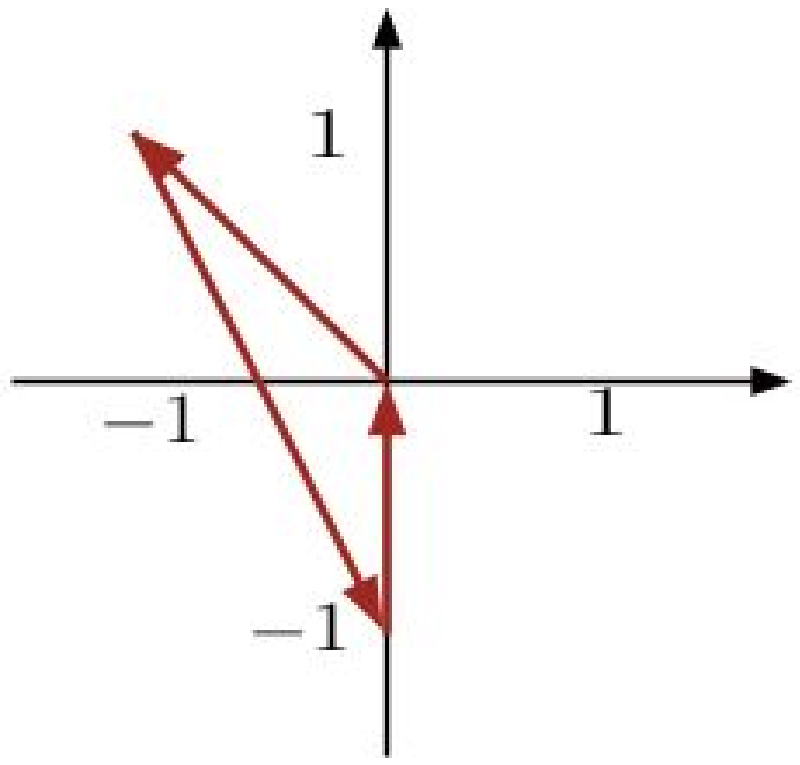}
\end{array}
\ee
In calculating the contribution of the first 
of the two paths we have to consider that
the steps out of the origin can start  
at any of the $(2t-2)$ intermediate steps 
\begin{widetext}
\bea
        a_0^{(2t)}(p\raw 0^+) &=& (2t-2) \left[ J_0(Kp)\right]^{2t-3}
        J_{-1}(Kp) e^{-i\phi} J_2((p+1)K) e^{-\sigma/2} 
        J_{-1}(Kp) e^{-i\phi} e^{-\sigma/2} a_0^{(0)}(p) \nonumber \\
        & \approx & (2t-2) \left(\frac{Kp}{2}\right)^2 
         e^{-2i\phi -\sigma} J_2(K) \, .
\eea
\end{widetext}
The same procedure leads to $\mathcal{P}_2= 
\mathcal{P}_1(\phi \raw -\phi)$.
We can now determine the correction to the 
diffusion coefficient:
\be
D=\frac{K^2}{4}\left[ 1-2 \cos (2 \phi) J_2(K)\right] +\mathcal{O}(\frac{1}{K}) \, ,
\label{c_12}
\ee
If we observe that we measured time in number of kicks, {\it i.e.} 
in units of $T/2$, \eq{c_12} exactly coincides with 
\eq{phase_effect}, in which time is measured in units of $T$.

\section{Full Counting Statistics of Cooper pair shuttling}
\label{levitov}

In this appendix we briefly review the technique of Full Counting Statistics 
(FCS) introduced in the seminal paper of Levitov and Lesovik~\cite{levitov96} 
and developed by Nazarov {\it et al.} (see Ref.~\onlinecite{kindermann03}).
We use such formalism to obtain \eqs{ale_fcs_general}{ale_fcs_general_1} presented in the text.
We present the general results for charge counting statistics 
without considering spin effects~\cite{lorenzo06}, which do not 
play any role in the case of Cooper pair transport analyzed in the paper.

Consider a conductor in which the charge dynamics
is determined by the Hamiltonian $\hat{H}_{\rm sys}$.
We are interested in the statistics of the current 
operator $\hat{I}$ through a given Section of the 
conductor.

We introduce the function $\fk{Z}(\Lambda,\lambda,\tau)
=\exp[-\fk{S}(\Lambda,\lambda,\tau)]$,
we will refer to as full counting statistics, formally defined through
\bea
        & &  \fk{Z}(\Lambda, \lambda, \tau) =
        \mbox{Tr} \underbrace{\left[\hat{\mathcal{U}}_{+\lambda}(\tau,0) \, \hat{\rho(0)} \,
        \hat{\mathcal{U}}^{\dag}_{-\lambda}(\tau,0) \right]}_{\hat{\rho} ^{(\lambda)} (\tau)} \, , 
\label{fcs_general} \\
        &  & \hat{\mathcal{U}}_{\pm \lambda}(\tau,0)=\overrightarrow{T} \exp^{-\imath \int_0^{\tau} \, 
        \hat{H}_{\Lambda \pm \lambda/2}(s)\, \frac{ds}{\hbar} } \\
        & &\hat{H}_{\Lambda \pm \lambda/2}=\hat{H}_{\trm sys}-
        \frac{\hbar}{e}\left(\Lambda \pm \lambda/2\right) \hat{I} \, .
\label{fcs_general_1}
\eea
It this expression the density matrix of the system is let evolve in time
according to two different  Hamiltonians for bra and ket. The trace of this ``modified''
 density matrix is the full counting statistics. The difference between the two Hamiltonians is 
due to the presence of $\lambda \neq 0$, 
that is called counting field.  It can be seen by direct verification that
$\fk{Z}(\Lambda,\lambda,\tau)$ is 
the generating function of current moments:
\be
\left. \partial^n_{\lambda} \fk{Z}(\Lambda,\lambda,\tau)
\right|_{\lambda, \Lambda=0}=
\frac{(- \imath)^n}{e^n} \int_{\left[ 0, \tau \right]^{\times n}} \! \! \! \! 
dt_1 \cdot dt_n \,
{\lavg  \hat{I}(t_1) \cdots \hat{I}(t_n) \ravg} \, .
\label{moments}
\ee
Let us also observe that the derivative of $\fk{S}(\Lambda,\lambda,\tau)$ gives us the 
cumulants of the current instead of its moments:
\bea
& & \partial^n_{\lambda} \left. \fk{S}(\Lambda, \lambda, \tau)\right|_{\lambda, \Lambda=0}
= -\partial^n_{\lambda} \left.
\left( \ln \fk{Z}(\Lambda, \lambda,\tau) \right)  \right|_{\lambda, \Lambda=0}=  \nonumber \\
& &  (- i/e)^n \int_{\left[ 0, \tau \right]^{\times n}} \! \! \! \! 
dt_1 \dots dt_n \,
{\lavg  \hat{I}(t_1) \dots \hat{I}(t_n) \ravg}_{\trm{conn.}} \, ,
\label{cumulant}
\eea
where $\avg{\cdot}_{\trm{conn.}}$ indicates the cumulant 
of the distribution function.
The FCS can be adapted to determine  
various integrated correlation functions  differing in 
the time ordering of the current operators and 
to obtain the correlation functions at arbitrary 
times rather than the integrated ones, thus revealing 
as a powerful tools in the investigation of 
the transport properties of a system.

Since the FCS can provide us information 
about the transport properties of the system, it 
would be meaningful to have an interpretation 
of $\fk{Z}(\Lambda,\lambda,\tau)$ directly in terms 
of charge transfer. 
As we are interested in the 
transport properties of a given system contacted 
to two electron reservoirs,\footnote{
The generalization to the multi-terminal case is 
straightforward.}
the properties of the system
are fully characterized by the knowledge
of the probability $P_{\tau}(N)$ of transferring 
$N$ charges in a given time interval $\tau$.
$P_{\tau}(N)$ is equivalently defined by its 
generating function $\chi(\lambda)=
\sum_N P_{\tau}(N) \exp(i N \lambda)$.
In particular all moments of the current distribution 
function can be obtained as
\begin{widetext}
\be
\int_{\left[ 0, \tau \right]^{\times m}} \! \! \! \! 
dt_1 \dots dt_m \,
{\lavg  \hat{I}(t_1) \dots \hat{I}(t_m) \ravg} 
=  \sum_N (eN)^m P_{\tau}(N) = 
(-i e)^m \partial^m_{\lambda} \left. 
\chi(\lambda) \right|_{\lambda=0} \, .
\label{moment_probab}
\ee
\end{widetext}
Due to the relation between the current 
moments and the function $\chi(\lambda)$, it would 
appear  natural to interpret the $\fk{Z}(\lambda, \Lambda,\tau)$
just constructed  as the 
generating function of the probabilities $P_{\tau}(N)$.
It can be shown that
\be
P_{\tau}(\Lambda,N)=\frac{1}{2\pi}\int_0^{2 \pi} d\lambda \,
\fk{Z}(\Lambda,\lambda,\tau) e^{-i q \lambda}
\ee
is a positive defined probability if $\fk{Z}(\lambda, \Lambda,\tau)$,
is independent on $\Lambda$, in which case
also $P_{\tau}(\Lambda,N)$ is.

Despite the previous analysis 
there are cases in which an interpretation 
of the transport properties in terms of classical 
probabilities is not possible. Coherent charge 
transfer between superconductors is a 
paradigmatic example.
The interpretation of  the FCS in this 
case has been discussed in literature 
(see Ref.~\onlinecite{nazarov_book02} and references therein).
Here we just notice that the appearance of 
non-positive or imaginary values 
of $P_{\tau}(\Lambda,N)$ are 
signature of the coherent nature of charge 
transfer between superconductors. 
 
We need to adapt the procedure to the case of a tunnel 
junction where the current operator is defined 
in terms of creation and annihilation operator 
in the two different leads.
Due to the linear coupling between the 
current and the counting field $\lambda$ 
in \eq{fcs_general_1}, charge conservation 
 ensures us that the counting field fulfills all properties 
 of a $U(1)$ gauge field.
In other words it means that the counting 
field plays the same role of the 
electromagnetic field, $\mbf{A}$.
In the case of a
tunnel junction between two electrons reservoirs
we are interested in the current through the region 
between the two leads.
The Hamiltonian is 
\bea
 & & \hat{H} = \sum_{A=L,R}  E_k \hat{c}_{A,k}^{\dag} \hat{c}_{A,k} + \hat{H}_T + \hat{H}_T^{\dag}\, ,
\label{tot_tun_h} \\
 & & \hat{H}_T = \sum_{k,h} T_{k,h} \hat{c}_{R,k}^{\dag} \hat{c}_{L,h}\, ,
\label{tunneling_h}
\eea
where $E_k$ is the $k$-dependent energy of electrons in the reservoirs, $T_{k,h}$ the tunneling amplitudes 
and $c_{L(R),k}$, $c_{L(R),k}^{\dag}$ are respectively the
annihilation and creation operator for electrons 
in the left (right) reservoir.
As the Hamiltonian is written in terms of 
operator defined at spatially separated points 
-that is at left ($L$) at right leads ($R$)-, 
we have to insert the counting field in the 
only way compatible with the $U(1)$ symmetry of the theory. 
It means that the counting field must 
enter through the Wilson line~\cite{peskin_book},
\be 
\exp \left( i \int_{L \raw R} \mbf{A} \cdot d\mbf{r} \right)=e^{- i(\Lambda+\lambda /2)} \, .
\label{wilson}
\ee
It results in a $\lambda$-dependent Hamiltonian $\hat{H}_{\Lambda + \lambda/2}$ obtained by the Hamiltonian in \eq{tot_tun_h} 
after the replacement 
\be
\hat{H}_T \rightarrow e^{-i (\Lambda + \lambda /2)} \hat{H}_T \, .
\label{cambio}
\ee
Indeed the integral in Eq.~(\ref{wilson}) depends on the 
specific path  going from the left to the right lead, 
but, in our case, the parameter $\Lambda \pm \lambda/2$ 
is defined in terms of the whole integral and, thus, the path-dependence in Eq.~(\ref{wilson}) is irrelevant.
We can now construct the FCS for the Cooper pair shuttle.

The model system we refer to is defined in Section 
\ref{sistema}.
For sake of simplicity, here we consider only 
the case $t_L=t_R=t_J$, $E_J^{(L)}=E_J^{(R)}=E_J$,
and $t_{\ra}=t_{\la}=t_C$. 
We consider the counting statistics for electrons 
 passing through a counter at left lead. 
The charge transfer in the Cooper pair shuttle
takes place through the Josephson effect, i.e. 
coherent tunneling of Cooper pairs.
According to the general procedure just described,  
we have to construct $\hat{H}_{\Lambda+\lambda/2}$
for the tunneling of Cooper pairs by modifying \eq{cambio}. 
In the Hamiltonian of the Cooper pair shuttle, \eq{kicked_rotator},
the tunneling term corresponds to the creation of a Cooper pair (of charge $2e$) into the grain, therefore 
the modification of the Hamiltonian in Eq~(\ref{cambio}) 
is given by
\be
\ket{n+1}\bra{n} \longrightarrow e^{- i (2 \Lambda + \lambda)} \ket{n+1}\bra{n} \, ,
\ee 
or equivalently
\be
\hat{H}_{\Lambda \pm \lambda/2} = \hat{H}_0(\phi_L \rightarrow \phi_L +2 \Lambda  \pm \lambda) \, .
\ee
In this way $\lambda$ is counting the charge in units of $e$.
Note that $\Lambda$ can be reabsorbed
into $\phi_L$ by the redefinition 
$\phi_L +\Lambda \raw \phi_L$.
Therefore, as long as the FCS depends on $\phi$, 
it will depend on $\Lambda$ and the interpretation 
in terms of probabilities that a certain number 
of Cooper pairs have traversed the shuttle
will be impossible.
The construction of the FCS including counting fields 
both at left and right contacts is straightforward and gives
\be
\hat{H}_{\lambda_L/2, \lambda_R/2} = \hat{H}_0(\phi_L \rightarrow \phi_L   
- \lambda_L \, ,\,\, \phi_R \rightarrow \phi_R   - \lambda_R) \, ,
\ee
as presented in \eq{ale_fcs_general_1}.


 \end{document}